\label{key}
\documentclass[journal,12pt, draftclsnofoot, onecolumn]{IEEEtran}
\pdfoutput=1
\usepackage{graphicx}
\usepackage{amsmath}
\usepackage{subfig}
\usepackage{mathtools}
\usepackage{epsfig}
\usepackage{float}
\usepackage{lipsum}
\usepackage{stfloats}
\usepackage{array}
\usepackage{amssymb}
\usepackage{cite}
\usepackage{url}
\usepackage{amsthm}
\usepackage{algorithm2e}
\usepackage{algorithmic}
\usepackage{multirow}
\usepackage{flushend}
\usepackage{upgreek}
\usepackage{dsfont}
\usepackage{bbm}
\usepackage{hhline}
\usepackage{textcomp}
\DeclarePairedDelimiter\ceil{\lceil}{\rceil}

\normalsize
\newcommand{\Phib}{\Phi_{\sf b}}
\newcommand{\lambdab}{\lambda_{\sf b}}
\newcommand{\Phiu}{\Phi_{\sf u}}

\newcommand{\lambdau}{\lambda_{\sf u}}

\newcommand{\shadSD}{\upsigma_{\scriptscriptstyle \rm dB}}

\newcommand{\mathcL}{\mathcal{L}}
\newcommand{\mathbE}{\mathbb{E}}
\newcommand{\mathbP}{\mathbb{P}}

\newcommand{\E}{\mathbb{E}}
\newcommand{\erf}{\mathrm{erf}}

\newcommand{\Na}{N_{\mathrm a}}

\newcommand{\imunit}{i}
\newcommand{\sinc}{\mathrm{sinc}}

\newcommand{\matI}{\boldsymbol{I}}
\newcommand{\matH}{\boldsymbol{h}}
\newcommand{\matF}{\boldsymbol{f}}
\newcommand{\vecx}{\boldsymbol{x}}

\newcommand{\sfk}{\mathsf{k}}

\newcommand{\SINR}{\mathsf{SINR}}

\newcommand{\SIR}{\mathsf{SIR}}

\newcommand{\SIRep}{\mathsf{SIR}^{\scriptscriptstyle \mathsf{Unif}}}
\newcommand{\SIRpa}{\mathsf{SIR}^{\scriptscriptstyle \mathsf{Eq}}}

\newcommand{\SINRep}{\mathsf{SINR}^{\scriptscriptstyle \mathsf{Unif}}}
\newcommand{\SINRpa}{\mathsf{SINR}^{\scriptscriptstyle \mathsf{Eq}}}

\newcommand{\bSE}{\bar{C}}
\newcommand{\SEep}{C^{\scriptscriptstyle \mathsf{Unif}}}
\newcommand{\bSEep}{\bar{C}^{\scriptscriptstyle \mathsf{Unif}}}
\newcommand{\SEpa}{C^{\scriptscriptstyle \mathsf{Eq}}}
\newcommand{\bSEpa}{\bar{C}^{\scriptscriptstyle \mathsf{Eq}}}

\newcommand{\sumSE}{C_{\scriptscriptstyle \Sigma}}
\newcommand{\bsumSE}{\bar{C}_{\scriptscriptstyle \Sigma}}

\newcommand{\bsumSEep}{\bar{C}_{\scriptscriptstyle \Sigma}^{\scriptscriptstyle \mathsf{Unif}}}

\newcommand{\bsumSEpa}{\bar{C}_{\scriptscriptstyle \Sigma}^{\scriptscriptstyle \mathsf{Eq}}}

\newcommand{\Lc}{L}

\newcommand{\Nc}{N_{\mathrm{c}}}

\newcommand{\noiseVar}{\sigma^2}

\newcommand{\Lref}{L_{\sf ref}}

\newcommand{\snrratio}{\varrho_{\sf \scriptscriptstyle SNR}}

\newtheorem{lem}{Lemma}
\newtheorem{prp}{Proposition}

\theoremstyle{definition}

\newtheorem{exmp}{Example}

\begin{document}
	
	\title{Massive MIMO Forward Link Analysis \\for Cellular Networks}
	
	\author{Geordie~George,~\IEEEmembership{Member,~IEEE},
		Angel~Lozano,~\IEEEmembership{Fellow,~IEEE}, \\ and Martin~Haenggi,~\IEEEmembership{Fellow,~IEEE}
\thanks{This work was supported by Project TEC2015-66228-P (MINECO/FEDER, UE), by the European Research Council under the H2020 Framework Programme/ERC grant agreement 694974, and by the U.S.~NSF through award CCF 1525904.

G. George was with the Department of Information and Communication Technologies, Universitat Pompeu Fabra (UPF), 08018 Barcelona, Spain. He is now with the Communication Systems Division, Fraunhofer Institute for Integrated Circuits IIS, Erlangen, Germany. E-mail: geordie.george@upf.edu.

A. Lozano is with the Department of Information and Communication Technologies, Universitat Pompeu Fabra (UPF), 08018 Barcelona, Spain. E-mail: angel.lozano@upf.edu.

Martin Haenggi is with the Department of Electrical Engineering, University of Notre Dame, Notre Dame, IN 46556, USA. E-mail: mhaenggi@nd.edu. 
}
}

\maketitle
	

\begin{abstract}
This paper presents analytical expressions for the signal-to-interference ratio (SIR) and the spectral efficiency in macrocellular networks with massive MIMO conjugate beamforming, both with a uniform and a channel-dependent power allocation. 
These expressions, which apply to very general network geometries, are asymptotic in the strength of the shadowing.
Through Monte-Carlo simulation, we verify their accuracy for relevant network topologies and shadowing strengths.
Also, since the analysis does not include pilot contamination, we further gauge through Monte-Carlo simulation the deviation that this phenomenon causes with respect to our results, and hence the scope of the analysis.
\end{abstract}

	
\section{Introduction}
	
	The tedious and time-consuming nature of system-level performance evaluations in wireless networks is exacerbated with massive MIMO \cite{5595728,6736761,marzetta_larsson_yang_ngo_2016,6798744,6375940,7031971,SIG-093,6415388,6843151,7880691,bjornson2018massive,8428665,Heath-Lozano-18}, as the number of active users per cell becomes hefty and the dimensionality of the channels grows very large.
	This reinforces the interest in analytical solutions, and such is the subject of this paper.
To embark upon the analysis of massive MIMO settings, we invoke tools from stochastic geometry that have been successfully applied already in non-MIMO \cite{Andrews-Baccelli-Ganti-coverage,StochGeoIntro-Haenggi-JSAC,MHaenggi12,7104127,Mukherjee-HetNet-JSAC12,ELSawy13,7243344}	
	and in MIMO contexts \cite{Ext_TWC,7448861,di2015stochastic,dhillon2013downlink}.
	
	We present expressions for the forward-link signal-to-interference ratio (SIR) and the spectral efficiency in macrocellular networks with massive MIMO conjugate beamforming, both with a uniform and a channel-dependent power allocation, and for different alternatives concerning the number of users per cell. The derived expressions apply to very general network geometries in the face of shadowing.
	These expressions allow:
	\begin{itemize}
		\item Testing and calibrating system-level simulators.
		\item Determining how many cells need to be simulated for some desired accuracy in terms of interference: the analysis subsumes an infinite field of cells whereas simulators necessarily feature a finite number thereof, which, if too small, results in a deficit of interference and in consequently optimistic performance predictions. 
		\item Gauging the impact of parameters such as the path loss exponent.
		\item Optimizing the number of active users as a function of the number of antennas and the path loss exponent.
		\item Assessing the benefits of a channel-dependent power allocation.
	\end{itemize}
	
	At the same time, the analysis is not without limitations, chiefly that pilot contamination is not accounted for. We explore this aspect by contrasting our analysis with simulation-based results that include the contamination, thereby delineating the scope of our results.
	
	The paper is organized as follows. The network and channel models are introduced in Section \ref{system model} and the forward-link conjugate beamforming SIRs, for uniform and equalizing power allocations, are derived in Section \ref{midterms}. Building on some preliminary analysis presented in Section \ref{sec:sys anal}, Sections \ref{TrumpSucks}--\ref{randomK} subsequently characterize the SIR distributions for fixed and Poisson-distributed numbers of users in each cell. The SIRs then lead to spectral efficiencies in Section \ref{SEsection}. The applicability of the results to relevant network geometries is illustrated in Section \ref{sec:validation} and the impact of noise and pilot contamination is assessed in Section \ref{tooth}. Finally, Section \ref{summary} concludes the paper.
	
	\section{Network Modeling}
	\label{system model}
	
	We consider a macrocellular network where each base station (BS) is equipped with $\Na \gg 1 $ antennas while users feature a single antenna.
	
	\subsection{Large-scale Modeling}
	\label{sec:large-scale}
	
The BS positions form a stationary and ergodic point process $\Phib \subset \mathbb{R}^2$ of density $\lambdab$, or a realization thereof, say a lattice network.
As a result,
the density of BSs within any region converges to $\lambdab > 0$ as this region's area grows \cite{7081363}.
In turn, the user positions conform to an independent point process $\Phiu \subset \mathbb{R}^2$ of density $\lambdau$, also stationary and ergodic.
Altogether, the models encompass virtually every macrocellular scenario of relevance.

Each user is served by the BS from which it has the strongest large-scale channel gain, 
and we denote by $K_\ell$ the number of users served by the $\ell$th BS. 

The large-scale channel gain includes path loss with exponent $\eta > 2$ and shadowing that is independent and identically distributed (IID) across the links. 
Specifically, the large-scale gain between the $\ell$th BS and the $k$th user served by the $l$th BS is
\begin{align}
G_{\ell,(l,k)} =  \frac{\Lref }{r_{\ell,(l,k)}^{\eta}} \, \chi_{\ell,(l,k)}   \qquad \ell,l \in \mathbb{N}_0, \, k \in \{0,\ldots,K_l - 1\},
\end{align} 	
with $\Lref$ the path loss intercept at a unit distance, $r_{\ell,(l,k)}$ the link distance, and $\chi_{\ell,(l,k)}$ the shadowing coefficient 
satisfying $\mathbb{E} \big[ \chi^{\delta} \big] <\infty$, where we have introduced $\delta = 2/\eta$. 
When expressed in dB, the shadowing coefficients have a standard deviation of $\shadSD$. In the absence of shadowing, $\chi=1$ and $\shadSD=0$.
		
Without loss of generality, we declare the $0$th BS as the focus of our interest and, for notational compactness, drop its index from the scripting. For the large-channel gains, for instance, this means that:
\begin{itemize}
\item	$G_{0,(l,k)} = G_{(l,k)}$ relates the BS of interest with the $k$th user served by the $l$th BS.
\item	$G_{\ell,(0,k)} = G_{\ell,(k)}$ relates the $\ell$th BS with the $k$th user served by the BS of interest.
\item $G_{0,(0,k)} = G_{(k)}$ relates the BS of interest with its own $k$th user.
\item $K_0 = K$ is the number of users served by the BS of interest.
\end{itemize}
The same scripting and compacting is applied to other quantities.

Let us consider an arbitrary user $k$ served by the BS of interest.
It is shown in \cite{7081363,keeler2014wireless,NRoss2016} that, as $\shadSD \rightarrow \infty$, irrespective of the actual BS positions, the \emph{propagation process} from that user's vantage, specified by $\{1/G_{\ell,(k)}\}_{\ell\in \mathbb{N}_0}$ and seen as a point process on $\mathbb{R}^+$, converges to
what the typical user---e.g., at the origin---would observe if the BS locations conformed to a homogeneous PPP on $\mathbb{R}^2$ with density $\lambdab$. 
 Moreover, by virtue of independent shadowing in each link, as $\shadSD \rightarrow \infty$ the propagation process from the vantage of each user becomes independent from those of the other users.
 Relying on these results, we embark on our analysis by regarding the propagation processes  
	\begin{align}	
	\left\{1/G_{\ell,(k)}\right\}_{\ell \in \mathbb{N}_0} \quad\quad k=0,\ldots,K-1 
	\end{align} 
	as IID and Poisson, anticipating that the results be applicable under relevant network geometries and realistic values of $\shadSD$; this is validated in Section \ref{sec:validation}.
	
	\subsection{Number of Users per BS}
	\label{rand num users}
	
	The modeling of $\{K_\ell\}$ 
	is a nontrivial issue.
	Even in a lattice network with equal-size cells, and let alone in irregular networks, disparities may arise across BSs because of the shadowing and the stochastic nature of the user locations. 
	
%
	
%
	
	
	In the absence of shadowing, the users served by a BS are those within its Voronoi cell, and their number---conditioned on the cell area---is a random variable with mean $\lambdau$ times the cell area; for instance, it is a Poisson random variable if $\Phiu$ is a PPP \cite{MHaenggi12}. Depending on how the cell area is distributed, then, the distribution of the number of users per cell can be computed. For instance, if $\Phiu$ is PPP, such number in a lattice network is Poisson-distributed with mean $\lambdau/\lambdab$ while, for an irregular network with PPP-distributed BSs, the corresponding distribution is computed approximately in \cite{7733098,6477048}.	
	
	With shadowing, a user need not be served by the BS in whose Voronoi cell it is located. 
	 Remarkably though, with strong and independent shadowing per link, a Poisson distribution with mean $\lambdau/\lambdab$ turns out to be a rather precise model---regardless of the BS locations---for the number of users per BS. This is because, as the shadowing strengthens, it comes to dominate over the path loss and, in the limit, all BSs become equally likely to be the serving one.
	Consider a region of area $\mathcal{A}$ having $\mathcal{A}  \lambdab$ BSs and $\mathcal{A} \lambdau$ users, all placed arbitrarily. As $\shadSD \rightarrow \infty$, the number of users served by each BS becomes binomially distributed, $B \Big( \mathcal{A} \lambdau, \frac{1}{\mathcal{A} \lambdab} \Big)$, because each user has equal probability $\frac{1}{\mathcal{A} \, \lambdab}$ of being served by any of the $\mathcal{A} \, \lambdab$ BSs. Now, letting $\mathcal{A} \rightarrow \infty$ while keeping $\lambdau/\lambdab$ constant, the binomial distribution converges to the Poisson distribution \cite{MHaenggi12}.
	
	In accordance with the foregoing reasoning, which is formalized in \cite{GLH18}, we model $\{K_\ell\}_{\ell \in \mathbb{N}_0}$ as IID Poisson random variables with mean $\mathbb{E}[K_\ell]=\bar{K} = \lambdau/\lambdab$.
	Noting that the unbounded tail of the Poisson distribution needs to be truncated at $\Na$, because such is the maximum number of users that can be linearly served by a BS with $\Na$ antennas, our analysis is conducted with Poisson $\{K_\ell\}_{\ell \in \mathbb{N}_0}$ and subsequently we verify the accuracy against simulations where the truncation is effected (see Examples \ref{ex10} and \ref{ex12}).
For the BS of interest specifically, we apply the Poisson PMF (probability mass function)
    	 \begin{align}
    	 \label{eq:Poisson PMF}
    	 f_K(\sfk) = \frac{\bar{K}^\sfk \, e^{-\bar{K}}}{\sfk!}  \qquad \quad \sfk \in \mathbb{N}_0 ,
    	 \end{align}
	 whose corresponding CDF (cumulative distribution function) is
    	 	\begin{equation}
    	 	\label{eq:Poisson CDF}
    	 	F_K(\sfk) = \frac{\Gamma(\sfk+1,\bar{K})}{\sfk!}  \qquad \quad \sfk \in \mathbb{N}_0 ,
    	 	\end{equation}
    	where $\Gamma(\cdot,\cdot)$ is the upper incomplete gamma function.


\subsection{Small-scale Modeling}
	
Let us focus on the local neighborhood of the $k$th user served by the BS of interest, wherein the large-scale gains $\{G_{\ell,(k)}\}_{\ell \in \mathbb{N}_0}$ apply. Without loss of generality, a system-level analysis can be conducted from the perspective of this user, which becomes the typical user in the network once we uncondition from $\{G_{\ell,(k)}\}_{\ell\in \mathbb{N}_0}$.
	
	Upon data transmission from the BSs, such $k$th user observes
	\begin{align}\label{eq:rec_sig0}
	y_{k} &= \sum_\ell \sqrt{G_{\ell,(k)}} \matH^*_{\ell,(k)} \, \vecx_\ell + v_{k}   \quad\qquad k = 0, \ldots, K - 1 ,
	\end{align}
	where 
	$\matH_{\ell,(k)} \sim \mathcal{N}_{\mathbb{C}}({\bf 0}, \matI)$ is the (normalized) reverse-link $\Na \times 1$ channel vector, $\matH^*_{\ell,(k)}$ is its forward-link reciprocal,
	 and $v_{k} \sim \mathcal{N}_{\mathbb{C}}(0,\noiseVar)$ is AWGN. The $\Na \times 1$ signal vector $\vecx_\ell$ emitted by the $\ell$th BS, intended for its $K_\ell$ users, satisfies $\E \big[ \| \vecx_\ell \|^2 \big] = P$ where $P$ is the per-base transmit power.
	
	\section{Conjugate Beamforming}
	\label{midterms}
\subsection{Transmit Signal}
	
	The signal transmitted by the $\ell$th BS is
	\begin{align}\label{eq:prec_sig}
	\vecx_\ell = \sum_{k=0}^{K_\ell - 1} \sqrt{\frac{P_{\ell,k}}{\Na}} \matF_{\ell,k} s_{\ell,k} ,
	\end{align}
	where $P_{\ell,k}$ is the power allocated to the data symbol $s_{\ell,k} \sim \mathcal{N}_{\mathbb{C}}(0,1)$, which is precoded by $\matF_{\ell,k}$ and intended for its $k$th user. 
	The power allocation satisfies
	\begin{align}\label{eq:powbud}
	\sum_{k=0}^{K_\ell-1} P_{\ell,k} = P
	\end{align}
	and, with conjugate beamforming and an average power constraint,
	\begin{align}
	\matF_{\ell,k} = \sqrt{\Na} \frac{\hat{\matH}_{\ell,(\ell,k)}}{\sqrt{\E \! \left[\|\hat{\matH}_{\ell,(\ell,k)}\|^2\right]}} \qquad\quad k = 0,\ldots,K_\ell - 1 ,
	\end{align}
	where $\hat{\matH}_{\ell,(\ell,0)}, \ldots, \hat{\matH}_{\ell,(\ell,K_\ell -1)}$ are the channel estimates gathered by the $\ell$th BS from the reverse-link pilots transmitted
	by its own users.
	
	Bringing (\ref{eq:rec_sig0}) and (\ref{eq:prec_sig}) together, the $k$th user served by the BS of interest observes
	\begin{align}\label{eq:rec_sig}
	y_{k} &= \sum_{\sfk=0}^{K-1} \sqrt{\frac{G_{(k)} P_{\sfk}}{\Na}} \matH^*_{(k)} \matF_{\sfk} s_{\sfk} + \sum_{\ell \neq 0} \sum_{\sfk=0}^{K_\ell -1}  \sqrt{\frac{G_{\ell,(k)} P_{\ell,\sfk}}{\Na}} \matH^*_{\ell,(k)} \matF_{\ell,\sfk} s_{\ell,\sfk} + v_{k} ,  
	\end{align}
whose first and second terms contain, respectively, the same-BS and other-BS transmissions.
	
	\subsection{SIR}
	
	We consider receivers reliant on channel hardening \cite{5595728}, whereby user $k$ served by the BS of interest regards $\mathbb{E} \big[ \matH_{(k)}^*\matF_k \big]$ as its precoded channel, with the expectation taken over the small-scale fading. The fluctuations of the actual precoded channel around this expectation constitute self-interference, such that (\ref{eq:rec_sig}) can be elaborated into
	\begin{align}
	y_k &= \underbrace{\sqrt{\frac{G_{(k)} P_k}{\Na}} \, \E \big[ \matH^*_{(k)} \matF_k \big] s_k}_{\mathsf{Desired \, signal}} + \underbrace{\sqrt{\frac{G_{(k)} P_k}{\Na}} \, \Big( \matH^*_{(k)} \matF_k - \E \big[\matH^*_{(k)} \matF_k \big] \Big) \, s_k}_{\mathsf{Self \text{-} interference}} \nonumber\\
	& \quad + \underbrace{\sum_{\sfk \neq k} \sqrt{\frac{G_{(k)} P_\sfk}{\Na}} \, \matH^*_{(k)} \matF_\sfk s_\sfk}_{\mathsf{Same \text{-} BS \, interference}} + \underbrace{\sum_{\ell \neq 0} \sum_{\sfk = 0}^{K_\ell -1} \sqrt{\frac{G_{\ell,(k)} P_{\ell,\sfk}}{\Na}} \, \matH^*_{\ell,(k)} \matF_{\ell,\sfk} s_{\ell,\sfk}}_{\mathsf{Other \text{-} BS \, interference}} + \underbrace{v_k}_{\mathsf{Noise}}
	\end{align}
	and the SINR is
	\begin{align}
	\SINR_k = \frac{G_{(k)} P_k \left| \E \big[ \matH_{(k)}^* \matF_k \big] \right|^2}{\mathrm{DEN}}
	\end{align}
	with
	\begin{align}
	\mathrm{DEN} &=  G_{(k)} P_k \, \mathrm{var} \big[ \matH_{(k)}^* \matF_k \big]
	+ \sum_{\sfk \neq k} G_{(k)} P_\sfk \, \E \! \left[ \big| \matH_{(k)}^* \matF_\sfk \big|^2\right] \nonumber \\
	& \quad + \sum_{\ell \neq 0} \sum_{\sfk=0}^{K_\ell-1} G_{\ell,(k)} P_{\ell,\sfk} \, \mathbb{E}\left[\left|\matH^*_{\ell,(k)} \matF_{\ell,\sfk}\right|^2\right] + \Na \, \noiseVar.
	\end{align}
	
	As indicated, the analysis in this paper ignores pilot contamination. It follows that, in interference-limited conditions ($\noiseVar/P \rightarrow 0$), the conjugate beamforming precoders at BS $\ell$ are
	\begin{align}
	\matF_{\ell,k} = \matH_{\ell,(\ell,k)} \qquad k = 0,\ldots, K_\ell - 1,
	\end{align}
	from which
	\begin{align}
	\left| \mathbb{E} \big[ \matH_{(k)}^* \matF_k \big] \right|^2 & = \left| \mathbb{E} \big[ \| \matH_{(k)} \|^2 \big] \right|^2 \\
	&  = \Na^2
	\end{align}
	and
	\begin{align}
	\mathrm{var} \big[ \matH_{(k)}^* \matF_k \big] & =  \mathbb{E} \big[ \| \matH_{(k)} \|^4 \big] - N^2_{\rm a} \\
	& = \Na \, (\Na+1) - N^2_{\rm a} \\
	& = \Na,
	\end{align}
	while, because of independence between user $k$ and other users $\sfk \neq k$, for both $\ell=0$ and $\ell > 0$ it holds that
	\begin{align}
	\mathbb{E} \! \left[\left|\matH^*_{\ell,(k)} \matF_{\ell,\sfk}\right|^2\right] 
	& = \Na .
	\end{align}
Altogether, in interference-limited conditions,
	\begin{align}
	\SIR_k &= \frac{\Na \, P_k \, G_{(k)} }{\sum_{\ell} \sum_{\sfk = 0}^{K_\ell -1} P_{\ell,\sfk} \, G_{\ell,(k)}}, 
	\end{align}
	which, invoking (\ref{eq:powbud}), further reduces to
	\begin{align}\label{eq:SIR0}
	\SIR_k &=  \frac{\Na \frac{P_k}{P} \,  G_{(k)}}{\sum_{\ell} G_{\ell,(k)} } . 
	\end{align}
	The foregoing ratio of channel gains can be seen to equal
	\begin{align}
	\frac{G_{(k)}}{\sum_{\ell} G_{\ell,(k)} } &= \frac{  G_{(k)}}{G_{(k)} + \sum_{\ell \neq 0} G_{\ell,(k)} } \\
	&= \frac{1}{1 + \frac{\sum_{\ell \neq 0} G_{\ell,(k)}}{G_{(k)}} } \\
	&= \frac{1}{1+1/\rho_k} \label{eq:varho1}
	\end{align}
	with
	\begin{align}
	\rho_k 
	&= \frac{G_{(k)}}{\sum_{\ell \neq 0} G_{\ell,(k)}} 
	\end{align}
	being the local-average SIR in single-user transmission \cite{Ext_TWC}. Hence, (\ref{eq:SIR0}) can be rewritten as
	\begin{align}
	\label{eq:SIR}
	\SIR_k 
	= \Na \, \frac{P_k/P}{1+1/\rho_k}. 
	\end{align}
	
	Two different power allocations, meaning two different formulations for $P_k/P$,
	are analyzed in this paper; the corresponding SIRs are presented next.
	
	\subsubsection{Uniform Power Allocation}
	
	With a uniform power allocation, $P_{k} = P / K$ and
	\begin{align}\label{eq:SIREP}
	\SIRep_k 
	&= \frac{\Na/K}{1+1/\rho_k} . 
	\end{align}
	
	\subsubsection{Equal-SIR Power Allocation}
	
	Alternatively, the SIRs can be equalized by setting \cite{6965818}
	\begin{align}
	P_{k}  &= P \, \frac{ \frac{\sum_{\ell} G_{\ell,(k)}}{G_{(k)}}}{\sum_{\sfk=0}^{K-1} \frac{\sum_{\ell} G_{\ell,(\sfk)}}{G_{(\sfk)}}} \\
	&= \frac{P \, (1+1/\rho_k)}{K + \sum_{\sfk=0}^{K-1} 1/\rho_\sfk} . 
	\end{align}
	Plugged into (\ref{eq:SIR}), this power allocation yields an SIR,
	common to all the users served by the BS of interest, of
	\begin{align}\label{eq:SIRPA}
	\SIRpa 
	= \frac{\Na/K}{1 + \frac{1}{K} \sum_{k=0}^{K-1} 1/\rho_k} . 
	\end{align}
	Introducing the harmonic mean of $\rho_0,\ldots,\rho_{K-1}$, namely
	\begin{align}
	\uprho_K 
	&= \frac{1}{\frac{1}{K} \sum_{k=0}^{K-1}1/\rho_k},\label{eq:uprho}
	\end{align}
	we can rewrite (\ref{eq:SIRPA}) as
	\begin{align}\label{eq:SIRPA2}
	\SIRpa 
	= \frac{\Na/K}{1+1/\uprho_K} .
	\end{align}
	
	Having formulated the SIRs for given large-scale link gains, 
	let us next characterize the system-level performance by releasing the conditioning on those gains.
	
	\begin{table}
		\caption{Parameter $s^\star$ for common values of the path loss exponent $\eta$ (the corresponding $\delta=2/\eta$ is also listed).}
		\label{tab:sstarvalues}
		\vspace{-0.1in}
		\begin{center}
			\begin{tabular}{ |c|c|c|}
				\hline
				$\eta$ & $\delta$ & $s^\star$  \\ \hline\hline
				3.5 & 0.571 & -0.672 \\ \hline
				3.6 & 0.556 & -0.71 \\ \hline
				3.7 & 0.540 & -0.747 \\ \hline
				3.8 & 0.526 & -0.783\\ \hline
			\end{tabular}
			\begin{tabular}{ |c|c|c|}
				\hline
				$\eta$ & $\delta$ & $s^\star$  \\ \hline\hline
				3.9 & 0.513  & -0.819\\ \hline
				4  & 0.5 & -0.854\\ \hline
				4.1 & 0.488 & -0.888\\ \hline
				4.2 & 0.476 & -0.922 \\ \hline
			\end{tabular}
		\end{center} 
		\vspace{-0.4cm}
	\end{table}
	
		\section{Spatial Distribution of $\rho_k$}
	\label{sec:sys anal}
	
	As the SIR formulations in (\ref{eq:SIREP}) and (\ref{eq:SIRPA}) indicate, a key ingredient is the distribution of $\rho_k = \frac{G_{(k)}}{\sum_{\ell \neq 0} G_{\ell,(k)}}$. 
	To characterize this distribution, we capitalize on results derived for the typical user in a PPP-distributed network of BSs \cite{Ext_TWC,7243344}, in accordance with the PPP convergence exposed in Section \ref{sec:large-scale}. Specifically, $\rho_0,\ldots,\rho_{K-1}$ are regarded as IID with CDF $F_\rho(\cdot)$, 
	where $\rho$ is the local-average SIR of the typical user in such a network \cite{Ext_TWC,7243344}.\footnote{Our analysis focuses on the typical cell, rather than the typical user (or, more precisely, the typical location). However, thanks to the Poisson convergence for strong shadowing, the propagation point process as seen from a user in the typical cell is statistically the same as that seen from the typical location in a shadowless PPP.}
	In the sequel, we employ a slightly simplified version of the result in \cite[Eq. 18]{Ext_TWC} as presented next. 
	
	\begin{lem}\cite{Ext_TWC}
	\label{ACdown}
	Define $s^\star<0$ as the solution---common values are listed in Table \ref{tab:sstarvalues}---to 
	\begin{align}
	\label{saul}
	{s^{\star}}^\delta \,\gamma(-\delta,s^\star)=0,
	\end{align}
	where $\gamma(\cdot,\cdot)$ is the lower incomplete gamma function. The CDF of $\rho$ satisfies
	\begin{align}
	\label{eq:SIRCDF_asymptotic1}
	\left\{ \begin{array}{l l}
	\!\! F_{\rho}(\theta) \simeq e^{s^\star/\theta} \qquad &  0 \leq \theta < \frac{1}{2+\epsilon} \\
	\!\! F_{\rho}(\theta) = 1 - \theta^{-\delta} \sinc \, \delta + B_{\delta} \! \left(\frac{\theta}{1-\theta}\right)  \qquad &   \frac{1}{2} \leq \theta < 1 \\
	\!\!F_{\rho}(\theta) = 1 - \theta^{-\delta} \sinc \, \delta \qquad &   \theta \geq 1,
	\end{array} \right.
	\end{align}
	where "$\simeq$" indicates asymptotic ($\theta \to 0$) equality while
	\begin{align}\label{bdelta}
	B_{\delta}(x) 
	&= \frac{ {}_2 F_1 \big(1, \delta+1; 2 \, \delta + 2; -1/x \big) \, \delta }{x^{1+2\,\delta} \;  \Gamma (2\,\delta + 2 ) \,  {\Gamma^2 (1-\delta)}}
	\end{align}
	with ${}_2 F_1$ the Gauss hypergeometric function. Setting
	\begin{equation}
	\epsilon = \frac{\log \! \left( 1-2^\delta \, \sinc \, \delta + B_{\delta}(1) \right)}{s^\star} - 2 ,
	\end{equation}
	we ensure $F_{\rho}\big(\frac{1}{2+\epsilon}\big)=F_\rho\big(\frac{1}{2}\big)$ and the CDF can be taken as constant therewithin.
	\end{lem}
	
	
	
As an alternative to the foregoing characterization, an exact but integral form can be obtained for $F_\rho(\cdot)$.

\begin{lem}\label{lem:charfunrho}
	\begin{align}
F_\rho(\theta)	&= \frac{1}{2} - \frac{1}{\pi} \int_{0}^{\infty} \Im \! \left \{\frac{e^{i(\theta+1)\omega}}{{}_1 F_1(1,1-\delta,i\theta\omega)}\right\} \frac{d\omega}{\omega} \label{eq:charfunrho}
	\end{align}
where $\Im \{ \cdot \}$ denotes imaginary part and ${}_1 F_1$ is the confluent hypergeometric or Kummer function.	
\begin{proof}
See Appendix \ref{app:proof of lemma1}.
\end{proof}
\end{lem}

While the results in the sequel are derived using either Lemma \ref{ACdown} or Lemma \ref{lem:charfunrho}, there is yet another useful alternative, namely computing $F_\rho(\cdot)$ via an approximate numerical inversion of the Laplace transform $\mathcal{L}_{1/\rho}(\mathsf{s}) = \mathbb{E}[e^{-\mathsf{s}/\rho}]$, presentation of which is relegated to Appendix \ref{app:NumLapInv}.
	
Equipped with the foregoing tools, let us proceed to characterize the spatial distribution of the SIRs in (\ref{eq:SIREP}) and (\ref{eq:SIRPA}). 

\section{Spatial SIR Distributions with Fixed $K$}
\label{TrumpSucks}

To begin with, let us characterize the SIR distributions for a fixed number of served users.
Besides having their own interest, the ensuing results serve as a stepping stone to their counterparts for $K$ conforming to the Poisson or to any other desired distribution.


\subsection{Uniform Power Allocation}

Starting from (\ref{eq:SIREP}), we can determine the CDF of $\SIRep_k$ as
\begin{align}
	F_{\SIRep_k}(\theta) 
	&= \mathbb{P} \! \left[\rho_k < \frac{\theta}{\Na/K - \theta} \right] \\
	&= \left\{ \begin{array}{l l} F_{\rho}\!\left(\frac{\theta}{\Na/K - \theta}\right) \qquad\quad & 0 < \theta < \Na/K \\
	1 \qquad\quad &  \theta \geq \Na/K,
	\end{array} \right. \label{eq:Fsirep}
\end{align}
which depends on $\Na$ and $K$ only through their ratio, $\Na/K$. 

Now, invoking Lemmas \ref{ACdown}--\ref{lem:charfunrho}, we can express the SIR distribution for the typical user in a massive MIMO network with fixed $K$ and a uniform power allocation.

\begin{prp}
\label{Honeymoon1}
The CDF of $\SIRep_k$ with a fixed $K$ satisfies
\begin{align}
\label{eq:CDFSIREP1}
\left\{ \begin{array}{l l}
\!\! F_{\SIRep_k}(\theta) \simeq e^{s^\star\left(\frac{\Na}{\theta \,K}-1\right)} \qquad &  0 \leq \theta  < \frac{\Na/K}{3 + \epsilon} \\
\!\! F_{\SIRep_k}(\theta) = 1 - \left(\frac{\Na}{\theta \,K}-1\right)^{\delta} \sinc \, \delta + B_{\delta} \! \left(\frac{\theta \,K}{\Na-2\,\theta \,K}\right)  \qquad &   \frac{\Na/K}{3 } \leq \theta < \frac{\Na / K}{2 } \\
\!\!F_{\SIRep_k}(\theta) = 1 - \left(\frac{\Na}{\theta \,K}-1\right)^{\delta} \sinc \, \delta \qquad &  \frac{\Na / K}{2} \leq \theta < \frac{\Na}{K} 
\end{array} \right.
\end{align}
with constant value within $\big[\frac{\Na/K}{3+\epsilon},\frac{\Na/K}{3}\big]$.
Alternatively, the CDF
can be computed exactly as
\begin{align}\label{eq:CDFSIREP2}
F_{\SIRep_k}(\theta) &= \frac{1}{2} - \frac{1}{\pi} \int_{0}^{\infty} \Im \! \left\{\frac{e^{\frac{i\omega}{1-\theta \, K/\Na}}}{{}_1 F_1\left(1,1-\delta,\frac{i\theta\omega}{\Na/K-\theta}\right)}\right\} \frac{d\omega}{\omega} \qquad  0 < \theta < \Na/K.
\end{align} 

\end{prp}

With either of these expressions, one can readily compute the percentage of users achieving a certain local-average performance for given $\Na/K$ and $\eta$. Furthermore, one can establish minima for $\Na/K$ given $\eta$ and given some target performance at a desired user percentile.

\begin{exmp}
\label{MontRavei}
Let $\eta=4$. In order to ensure that no more than $3\%$ of users experience an SIR below $0$ dB, it is required that $\Na/K \geq 5$.
\end{exmp}

\subsection{Equal-SIR Power Allocation}

From 
(\ref{eq:SIRPA2}), the CDF of $\SIRpa$ can be expressed as
\begin{align}
F_{\SIRpa}(\theta) 
&= \mathbb{P} \! \left[\uprho_K < \frac{\theta}{\Na/K - \theta} \right] \\
&= \left\{ \begin{array}{l l} F_{\uprho_K}\!\left(\frac{\theta}{\Na/K - \theta}\right) \qquad\quad & 0 < \theta < \Na/K \\
1 \qquad\quad &  \theta \geq \Na/K
\end{array} \right.
\end{align}
where, recall, $\uprho_K$ is the harmonic mean of $\rho_0,\ldots,\rho_{K-1}$. 

While an explicit expression such as (\ref{eq:CDFSIREP1}) seems difficult to obtain for $F_{\SIRpa}(\cdot)$,
an integral form similar to (\ref{eq:CDFSIREP2}) is forthcoming.

\begin{prp}
\label{prp:SIREq_fixedK}
	The CDF of $\SIRpa$ with a fixed $K$ is
	\begin{align}
			F_{\SIRpa}(\theta) = \frac{1}{2} - \frac{1}{\pi} \int_{0}^{\infty} \Im \! \left\{\frac{e^{i\omega}}{\left\{{}_1 F_1\!\left(1,1-\delta,i\theta\omega/\Na\right)\right\}^K}\right\} \frac{d\omega}{\omega} \qquad 0 < \theta < \Na/K.  \label{eq:CDFSIRPC1}
	\end{align}
	\begin{proof}
		See Appendix \ref{app:SIREq_fixedK}.
	\end{proof}
\end{prp}

What can be established explicitly is that, as $\Na, K \rightarrow \infty$ with ratio $\Na/K$,
\begin{align}\label{eq:assymsirPC}
\SIRpa &\rightarrow
\frac{\Na}{K} \, (1-\delta) 
\end{align}
which follows from
\begin{align}
	\lim\limits_{K \rightarrow \infty} \frac{1}{K} \sum_{k=0}^{K-1} \frac{1}{\rho_k} 
	&= \mathbb{E} \big[ 1/\rho \big] \\
	&= \frac{\delta}{1 - \delta}, \label{eq:MISR}
\end{align}
where (\ref{eq:MISR}) is established in \cite{6897962}.
\begin{exmp}\label{ex1}
	\begin{figure} 
		\centering
		\includegraphics [width = 0.65\columnwidth]{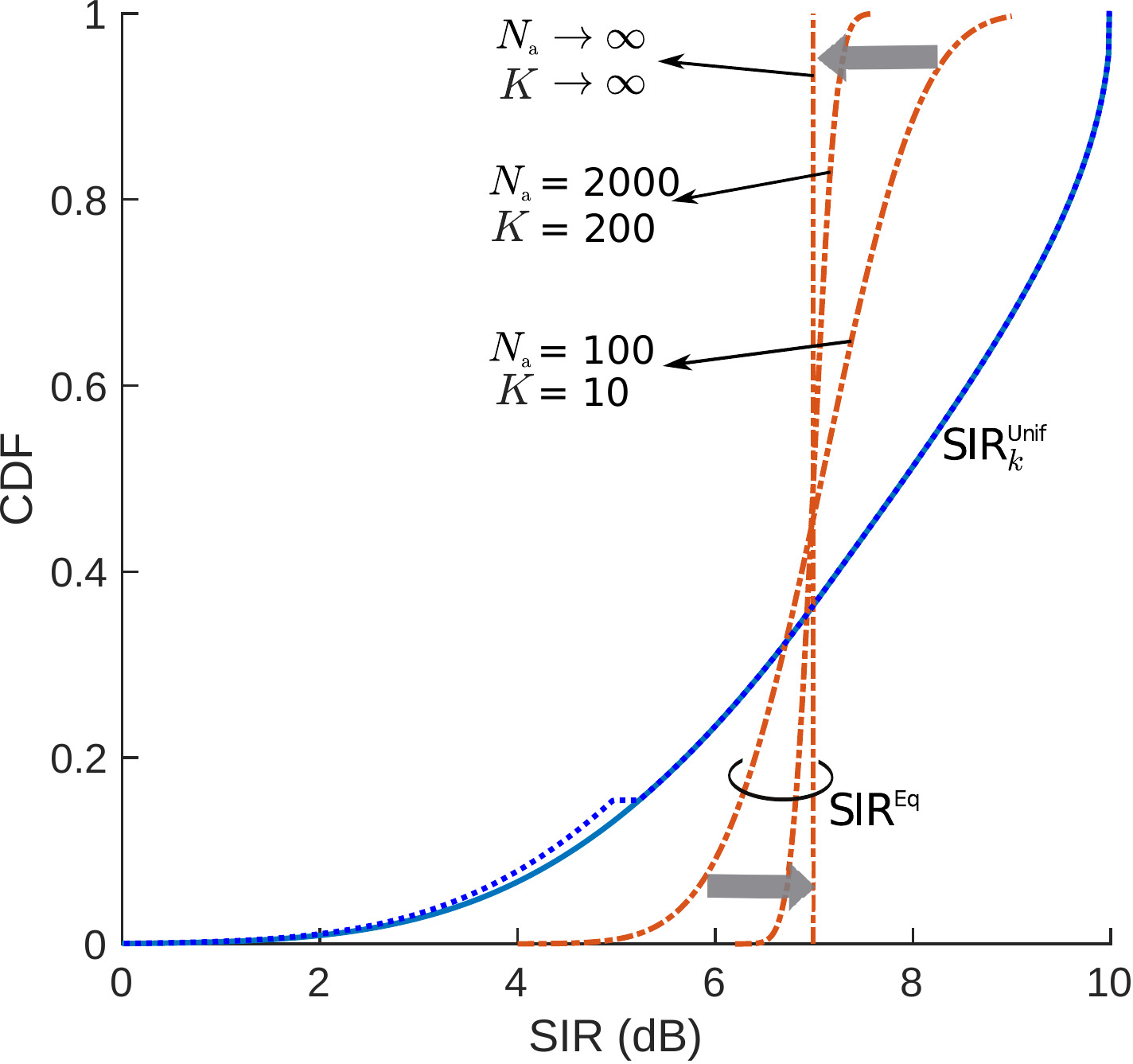} 
		\caption{CDFs of $\SIRep_k$ and $\SIRpa$ for $\Na/K = 10$ and $\eta = 4$.
		For $\SIRep_k$, the dotted and solid curves correspond, respectively, to (\ref{eq:CDFSIREP1}) and (\ref{eq:CDFSIREP2}) in Prop. \ref{Honeymoon1}. The dashed curves for $\SIRpa$ are obtained via Prop. \ref{prp:SIREq_fixedK}, with the block arrows indicating the convergence for $\Na,K \to \infty $ as established in (\ref{eq:assymsirPC}).}
		\label{fig:figure1}
	\end{figure}
$F_{\SIRpa}(\cdot)$ and $F_{\SIRep_k}(\cdot)$ are compared in Fig. \ref{fig:figure1}, for $\Na/K = 10$ and  $\eta = 4$. For $\SIRep_k$, the dotted and solid curves correspond, respectively, to (\ref{eq:CDFSIREP1}) and (\ref{eq:CDFSIREP2}) in Prop. \ref{Honeymoon1}. The dashed curves for $\SIRpa$, obtained 
via Prop. \ref{prp:SIREq_fixedK}, can be observed to converge, as per (\ref{eq:assymsirPC}), to
\begin{align}
\frac{\Na}{K} \, (1-\delta) & = 10 \, (1-1/2) \\
& = 7 \, \mathrm{dB}
\end{align}
 as $\Na, K \rightarrow \infty$.
\end{exmp}

Note from the foregoing example that, for $\Na = 100$ and $K = 10$ onwards, over 90\% of users have an $\SIRpa$ within $ 1$ dB of its spatial average. Besides confirming the effectiveness of the simple SIR-equalizing power allocation for massive MIMO, this allows establishing $\Na/K$ rather accurately by directly equating $\frac{\Na}{K} (1-\delta)$ to the desired SIR.


\section{Spatial SIR Distribution with Poisson-distributed $K$}
\label{randomK}

Let us now allow $K$ to adopt a Poisson distribution, which, as argued, captures well the variabilities caused by network irregularities and shadowing. Similar derivations could be applied to other distributions if, for instance, one wishes to further incorporate activity factors for the users or channel assignment mechanisms.



\subsection{Uniform Power Allocation}

The expectation of (\ref{eq:CDFSIREP1}) in Prop. \ref{Honeymoon1} over (\ref{eq:Poisson PMF}) yields the following result.

\begin{prp}\label{prp:SIRUnif_randK}
	The CDF of $\SIRep_k$ with a Poisson-distributed $K$ satisfies
	\begin{align}
	F_{\SIRep_k}(\theta) &\approx \frac{1}{e^{\bar{K}}-1} \sum_{\sfk=1}^{\ceil{\frac{\Na/\theta}{3+\epsilon}}-1}e^{s^\star\left(\frac{\Na}{\theta \,\sfk}-1\right)} \frac{\bar{K}^\sfk}{\sfk!} + \frac{1}{e^{\bar{K}}-1} \sum_{\sfk=\ceil{\frac{\Na/\theta}{3+\epsilon}}}^{\ceil{\frac{\Na}{3\theta}}-1} e^{s^\star(\epsilon+2)} \,  \frac{\bar{K}^\sfk}{\sfk!} 
	\nonumber\\
	& \quad + \frac{1}{e^{\bar{K}}-1} \sum_{\sfk=\ceil{\frac{\Na}{3\theta}}}^{\ceil{\frac{\Na}{2\theta}}-1} \left[ 1 - \left(\frac{\Na}{\theta \,\sfk}-1\right)^{\delta} \sinc \, \delta + B_{\delta} \! \left(\frac{\theta \,\sfk}{\Na-2\,\theta \,\sfk}\right)\right] \frac{\bar{K}^\sfk}{\sfk!}\nonumber   \\
	& \quad + \frac{1}{e^{\bar{K}}-1} \sum_{\sfk=\ceil{\frac{\Na}{2\theta}}}^{\ceil{\Na/\theta}-1} \left[ 1 - \left(\frac{\Na}{\theta \,\sfk}-1\right)^{\delta} \sinc \, \delta \right] \frac{\bar{K}^\sfk}{\sfk!} + \frac{ 1 -\Gamma\left(\ceil{\frac{\Na}{\theta}},\bar{K}\right)}{\left(1 - e^{-\bar{K}}\right)  \left(\ceil{\frac{\Na}{\theta}}-1\right)!} \label{eq:CDFSIREP_randU0}
	\end{align}	
	where $B_\delta(\cdot)$ is as per (\ref{bdelta}).
	Exactly,
	\begin{align}\label{eq:CDFSIREP_randU}
F_{\SIRep_k}(\theta) &= \frac{1}{2} - \frac{1}{\pi \, (e^{\bar{K}}-1)}  \sum_{\sfk=1}^{\ceil{\Na/\theta}-1} \frac{\bar{K}^\sfk}{\sfk!} \int_{0}^{\infty} \Im\left\{\frac{e^{\frac{i\omega}{1-\theta \, \sfk/\Na}}}{{}_1 F_1\left(1,1-\delta,\frac{i\theta\omega}{\Na/\sfk-\theta}\right)}\right\} \frac{d\omega}{\omega} \nonumber \\
&\qquad\qquad\qquad \qquad \qquad\qquad\quad + \frac{ 1 -\Gamma\left(\ceil{\frac{\Na}{\theta}},\bar{K}\right)}{2\left(1 - e^{-\bar{K}}\right)  \left(\ceil{\frac{\Na}{\theta}}-1\right)!}.
\end{align}
	
	\begin{proof}
		See Appendix \ref{app:SIRUnif_randK}.
	\end{proof}
\end{prp}

For $\Na, \bar{K} \rightarrow \infty$ with fixed $\Na/\bar{K}$, it can be verified that 
$\Na/K \rightarrow \Na/\bar{K}$ with convergence in the mean-square sense, and therefore in probability.
As a consequence, $\SIRep$, which depends on $\Na$ and $K$ only through their ratio, progressively behaves as if this ratio were fixed at $\Na/\bar{K}$ despite the Poisson nature of $K$. This behavior is clearly in display in Fig. \ref{fig:figure11}, where $F_{\SIRep_k}(\cdot)$ can be seen to approach its value for fixed $K=\bar{K}$.


\subsection{Equal-SIR Power Allocation}

In this case, we expect the expression in Prop. \ref{prp:SIREq_fixedK} over (\ref{eq:Poisson PMF}) to obtain the following.

\begin{prp}
\label{prp:SIREq_randK}

Since the user's SIR is a valid quantity only for $K\geq1$, i.e., when the BS has at least one user, $F_{\SIRpa}(\cdot)$ with Poisson-distributed $K$ equals
\begin{align}
F_{\SIRpa}(\theta) &=  \sum_{\sfk=1}^{\infty} F_{\SIRpa|K=\sfk}(\theta) \, \frac{f_K(\sfk)}{1-F_K(0)} \\
&= \frac{1}{2} - \frac{1}{\pi (e^{\bar{K}}-1)} \! \int_{0}^{\infty}\!\! \Im\left\{e^{i\omega}  \left[e^{\frac{\bar{K}}{ {}_1 F_1\left(1,1-\delta,\frac{i\theta\omega}{\Na}\right)}}-1\right] \right\}\! \frac{d\omega}{\omega} \label{eq:CDFSIRPA_randU},
\end{align}
where we recalled the expression for $F_{\SIRpa|K=\sfk}(\cdot)$ from (\ref{eq:CDFSIRPC1}) and invoked (\ref{eq:Poisson PMF})--(\ref{eq:Poisson CDF}). 
\end{prp}

Again, because of the convergence $\Na/K \to \Na/\bar{K}$ for $\Na,\bar{K} \to \infty$, $\SIRpa$ hardens to its fixed-$K$ value with $K = \bar{K}$. In this case, this corresponds to $\Na \, (1-\delta)/\bar{K}$ as demonstrated in Fig. \ref{fig:figure11}.

\begin{exmp}\label{examrandK}
		\begin{figure} 
			\centering
			\includegraphics [width = 0.65\columnwidth]{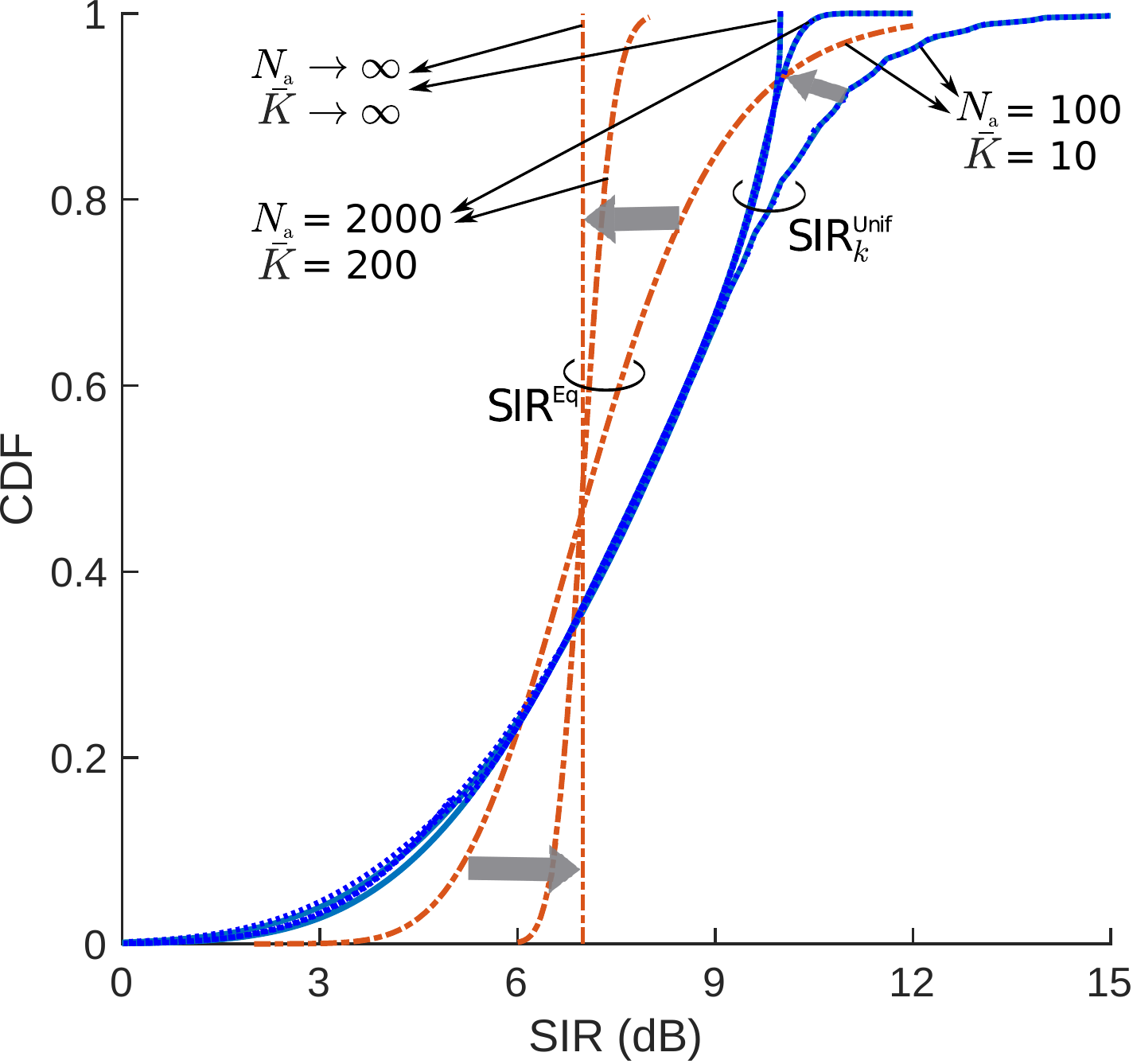} 
			\caption{CDFs of $\SIRep_k$ and $\SIRpa$ with $\Na/\bar{K} = 10$ and $\eta = 4$.
			For $\SIRep_k$, the dotted and solid lines correspond to Prop. \ref{prp:SIRUnif_randK}, respectively (\ref{eq:CDFSIREP_randU0}) and (\ref{eq:CDFSIREP_randU}). The dashed curves for $\SIRpa$ correspond to Prop. \ref{prp:SIREq_randK}.
			The block arrows indicate the evolution for $\Na,\bar{K} \rightarrow \infty$.}
			\label{fig:figure11}
		\end{figure}
	$F_{\SIRpa}(\cdot)$ and $F_{\SIRep}(\cdot)$ with a Poisson-distributed $K$ are compared in Fig. \ref{fig:figure11}, for $\Na/\bar{K} = 10$ and $\eta = 4$. 
	For $\SIRep_k$, the dotted and solid lines correspond to Prop. \ref{prp:SIRUnif_randK}, respectively (\ref{eq:CDFSIREP_randU0}) and (\ref{eq:CDFSIREP_randU}). The dashed curves for $\SIRpa$ correspond to Prop. \ref{prp:SIREq_randK}. 
\end{exmp}

\section{Spectral Efficiency}
\label{SEsection}

With each user's SIR stable over its local neighborhood thanks to the channel hardening, the
spectral efficiency of the typical user is directly 
\begin{align}
	C_k = \log_2(1+\SIR_k), 
\end{align}
and the spatial distribution thereof can be readily obtained as
\begin{align}
	F_{C_k}(\zeta) &= \mathbb{P} \big[ \log_2(1+\SIR_k) < \zeta \big] \\
	&= 	F_{\SIR_k}(2^\zeta-1)\label{eq:secdf}
\end{align}
for each setting considered in the foregoing sections.


\begin{exmp}\label{exmp:5pertSEfix}
	Let $\eta = 4$, $\Na = 100$ and $K = 10$. The $3\%$-ile user spectral efficiency, computed by numerically solving for $\zeta$ in $F_{\SIRep_k}(2^\zeta-1) = 0.03$ and $F_{\SIRpa}(2^\zeta-1) = 0.03$, respectively under uniform and equal-SIR power allocations, are $1.6$ b/s/Hz and $2.2$ b/s/Hz.
\end{exmp}

\begin{exmp}\label{exmp:5pertSErand}
Reconsidering Example \ref{exmp:5pertSEfix}, but with $K$ being Poisson with mean $\bar{K} = 10$, the $3\%$-ile user spectral efficiencies, respectively under uniform and equal-SIR power allocations, are $1.51$ b/s/Hz and $1.94$ b/s/Hz.
\end{exmp}

From its user spectral efficiencies, the sum spectral efficiency at the BS of interest can be found as
\begin{align}\label{eq:sumSE}
C_{\Sigma} = \sum_{k=0}^{K-1}C_k , 
\end{align}
which is zero whenever the BS serves no users.

Explicit expressions for the spatial averages of the user and the sum spectral efficiencies, $\bSE = \mathbb{E} \big[ C_k \big]$ and $\bsumSE = \mathbb{E} \! \left[\sumSE\right]$ with expectation over all possible propagation processes, are presented next.

\subsection{Average Spectral Efficiency: Uniform Power Allocation}


\begin{prp}
\label{Esterilla}
With a uniform power allocation and a fixed $K$, the spatially averaged user spectral efficiency equals
\begin{align}
\label{Tertulia}
\bSEep =  \log_2(e) \,\int_{0}^{\infty} \frac{ 1-e^{-z \Na/K}}{ {}_1 F_1 \big( 1,1-\delta,z \big)} \, \frac{{d}z}{z}
\end{align}
and the spatially averaged sum spectral efficiency is $\bsumSEep = K \bSEep$.

\begin{proof}
	See Appendix \ref{Messi}.
\end{proof}
\end{prp}

For the special case of $\eta=4$, i.e., for $\delta=1/2$, Prop. \ref{Esterilla} reduces to
\begin{align}
\bSEep &=  \log_2(e) \,\int_{0}^{\infty} \frac{ 1-e^{-z \Na/K}}{1 + e^z \sqrt{\pi z} \, \erf\sqrt{z}} \, \frac{{d}z}{z}
\label{ACkaputt}
\end{align}
thanks to ${}_1 F_1 \big( 1,1/2,z \big) \equiv 1 + e^z \sqrt{\pi z} \, \erf\sqrt{z}$, an equivalence that is further applicable in the characterizations that follow.

Since the denominator of the integral's argument in (\ref{ACkaputt}), and more generally in (\ref{Tertulia}), is strictly positive for $z \geq 0$, we can deduce by inspection that $\bSEep$ shrinks if we increase $K$ with a fixed $\Na$, i.e., if we add more users with a fixed number of antennas. However, this reduction is sublinear in $K$ and thus $\bsumSEep$ grows as we add more users. Thus, $K$ should be set to the largest possible value that ensures an acceptable performance for the individual users---recall Example \ref{MontRavei}---and the corresponding average performance for both individual users and cells can then be readily computed by means of (\ref{Tertulia}) or (\ref{ACkaputt}).

\begin{exmp}
Example \ref{MontRavei} established that, for $\eta=4$, we should have $\Na/K \geq 5$ to ensure that no more than $3\%$ of users fall below $0$ dB of SIR. Plugged into Prop. \ref{Esterilla}, such $\Na/K$ returns $\bSEep=1.97$ b/s/Hz while the corresponding sum spectral efficiency becomes $\bsumSEep=\Na \, \bSEep/5 = 0.39 \, \Na$.
\end{exmp}


Expecting the spatially averaged user spectral efficiency in (\ref{Tertulia}) over $K$, via the Poisson distribution in (\ref{eq:Poisson PMF}), we obtain
\begin{align}
	\bSEep &= \sum_{\sfk=1}^{\infty} \bSEep|_{K=\sfk} \, \frac{f_K(\sfk)}{1-F_{K}(0)} \\
	&= \log_2(e) \,\int_{0}^{\infty} \frac{ 1- \frac{1}{e^{\bar{K}}-1} \sum_{\sfk=1}^{\infty} e^{-z \frac{\Na}{\sfk}} \bar{K}^\sfk/\sfk!}{{}_1 F_1 \big( 1,1-\delta,z \big)} \, \frac{{d}z}{z}.
	\label{eq:avSEeprandU}
\end{align}
The corresponding spatial average of the sum spectral efficiency, recalling (\ref{eq:sumSE}), is
\begin{align}
	\bsumSEep &= \sum_{\sfk=1}^{\infty} \sfk \, \bSEep|_{K=\sfk} \, f_K(\sfk) \\
	&=  \frac{\log_2 e}{e^{\bar{K}}} \,\int_{0}^{\infty} \frac{\sum_{\sfk=1}^{\infty} \sfk \left(1-e^{-z \Na/\sfk}\right)  \bar{K}^\sfk/\sfk!}{{}_1 F_1 \big( 1,1-\delta,z \big)} \, \frac{{d}z}{z} . \label{eq:avcellSEerandU}
\end{align}

\begin{exmp}\label{exmp:avgSEbeg}
	Let $\eta = 4$ and $\Na = 100$ with a uniform power allocation.
For $K=10$, (\ref{ACkaputt}) returns an average user spectral efficiency of $\bSEep = 2.76$ b/s/Hz and an average sum spectral efficiency of $\bsumSEep = 10 \, \bSEep= 27.6$ b/s/Hz per BS.
For  $K$ being Poisson with mean $\bar{K} = 10$, (\ref{eq:avSEeprandU}) returns $\bSEep = 2.84$ b/s/Hz while (\ref{eq:avcellSEerandU}) yields $\bsumSEep = 27.08$ b/s/Hz per BS.
\end{exmp}

\subsection{Average Spectral Efficiency: Equal-SIR Power Allocation}


\begin{prp}
\label{Esterilla2}
With an equal-SIR power allocation and a fixed $K$, the spatially averaged user spectral efficiency equals
\begin{align}
\label{Tertulia2}
\bSEpa =  \log_2(e) \,\int_{0}^{\infty} \frac{ 1-e^{-z}}{  {}_1 F_1\left(1,1-\delta,z/\Na \right)^K} \, \frac{{d}z}{z},
\end{align}
and the spatially averaged sum spectral efficiency is  $\bsumSEpa = K \bSEpa$.

\begin{proof}
	See Appendix \ref{Messi}.
\end{proof}
\end{prp}

For $\Na, K \rightarrow \infty$, recalling how the SIR hardens to $(1-\delta) \, \Na/K$, we have that $\bSEpa$ hardens to
$
\log_2 \! \left[1+ (1-\delta) \, \frac{\Na}{K} \right].
$


Expecting (\ref{Tertulia2}) over $K$, via (\ref{eq:Poisson PMF}), gives
\begin{align}
\bSEpa 
&=\frac{\log_2 e}{1-F_K(0)} \,\int_{0}^{\infty} \frac{1-e^{-z }}{z} \sum_{\sfk=1}^{\infty} \frac{f_K(\sfk)}{ {}_1 F_1\left(1,1-\delta,z/\Na \right)^\sfk} \, {d}z \\
&= \frac{\log_2 e}{e^{\bar{K}}-1} \,\int_{0}^{\infty} \frac{ 1-e^{-z }}{z} \, \left(e^{\frac{\bar{K}}{{}_1 F_1\left(1,1-\delta,z/\Na \right)}}-1\right) \, {d}z.
 \label{eq:bSEparandU}
\end{align}
In turn, the average sum spectral efficiency with expectation over (\ref{eq:Poisson PMF}) becomes
\begin{align}
\bsumSEpa 
&= \log_2(e) \,\int_{0}^{\infty} \frac{1-e^{-z }}{z}  \sum_{\sfk=1}^{\infty} \frac{ \sfk \, f_K(\sfk) }{  {}_1 F_1\left(1,1-\delta,z/\Na \right)^\sfk}  \, {d}z \\
&= \frac{\log_2 e}{e^{\bar{K}}} \,\int_{0}^{\infty} \frac{1-e^{-z }}{z} \frac{\bar{K} \, e^{\frac{\bar{K}}{{}_1 F_1\left(1,1-\delta,z/\Na \right)}}}{{}_1 F_1\left(1,1-\delta,z/\Na \right)}  \, {d}z. \label{eq:avcellSEerandU2}
\end{align}

\begin{exmp}\label{exmp:avgSEfin}
	Let $\eta = 4$ and $\Na = 100$ with an equal-SIR power allocation.
For $K=10$, (\ref{Tertulia2}) returns an average user spectral efficiency of $\bSEpa = 2.61$ b/s/Hz	and an average sum spectral efficiency of $\bsumSEpa = 10 \, \bSEpa= 26.1$ b/s/Hz per BS.
For $K$ being Poisson with mean $\bar{K} = 10$, (\ref{eq:bSEparandU}) returns $\bSEpa = 2.69$ b/s/Hz while  (\ref{eq:avcellSEerandU2}) yields $\bsumSEpa = 25.56$ b/s/Hz per BS.
\end{exmp}


\begin{table}
	\centering
	\caption{Spectral efficiency (b/s/Hz) with $\Na = 100$ and $\eta =4$.}
	\label{tab:SEtable}
	\vspace{-0.1in}
	\begin{center}
		\begin{tabular}{|c|c||c|c|}
			\hline
			\multicolumn{2}{|c||}{} & $K = 10$ & Poisson $K$ with $\bar{K}=10$ \\  \hline\hline
			\multirow{2}{*}{$3\%$-ile b/s/Hz per user} & $\SEep$  & 1.60 & 1.51 \\ \cline{2-4} 
			& $\SEpa$   & 2.20 & 1.94 \\ \hline
			\multirow{2}{*}{Average b/s/Hz per user}  & $\bSEep$ & 2.76 & 2.84 \\ \cline{2-4} 
			& $\bSEpa$  & 2.61 & 2.69 \\ \hline
			\multirow{2}{*}{Average b/s/Hz per BS}  & $\bsumSEep$  & 27.6 & 27.08 \\ \cline{2-4} 
			& $\bsumSEpa$  & 26.1 & 25.56 \\ \hline
		\end{tabular}
	\end{center} 
\end{table}

For the reader's convenience, the results from Examples \ref{exmp:avgSEbeg}--\ref{exmp:avgSEfin} are brought together in Table \ref{tab:SEtable}, alongside the $3\%$-ile user spectral efficiencies from Examples \ref{exmp:5pertSEfix}--\ref{exmp:5pertSErand}. Not surprisingly, with an equal-SIR power allocation, the $3\%$-ile user spectral efficiency improves significantly, but the spatial average slips relative to a uniform power allocation. These results also corroborate the observation \cite[Remark 4.1]{6965818} that the sum spectral efficiency with a uniform power allocation is higher than with an equal-SIR power allocation. Fairness does come at a price.


\section{Application to Relevant Network Geometries}
\label{sec:validation}

Let us proceed to verify, via Monte-Carlo, that the analytical results in the foregoing sections closely abstract the performance of network geometries of interest with practical values for the shadowing standard deviation. 
The number of network snapshots is chosen to ensure a $95$\% confidence interval of $\pm 0.07\%$ (absolute value) in the CDFs, which at the median corresponds to at most $\pm 0.06$ dB in SIR.


\subsection{Hexagonal Lattice Networks}

Let the BS locations $\Phib$ conform to a hexagonal lattice. In the following examples, Monte-Carlo results are generated for the users served by the BS at the center of a lattice of 499 hexagonal cells. The shadowing is lognormal.



\begin{exmp}
%
\begin{figure}[!tbp]
	\centering
	\subfloat[Uniform power allocation; analysis from Prop. \ref{Honeymoon1}.]{\includegraphics[width=0.43\linewidth]{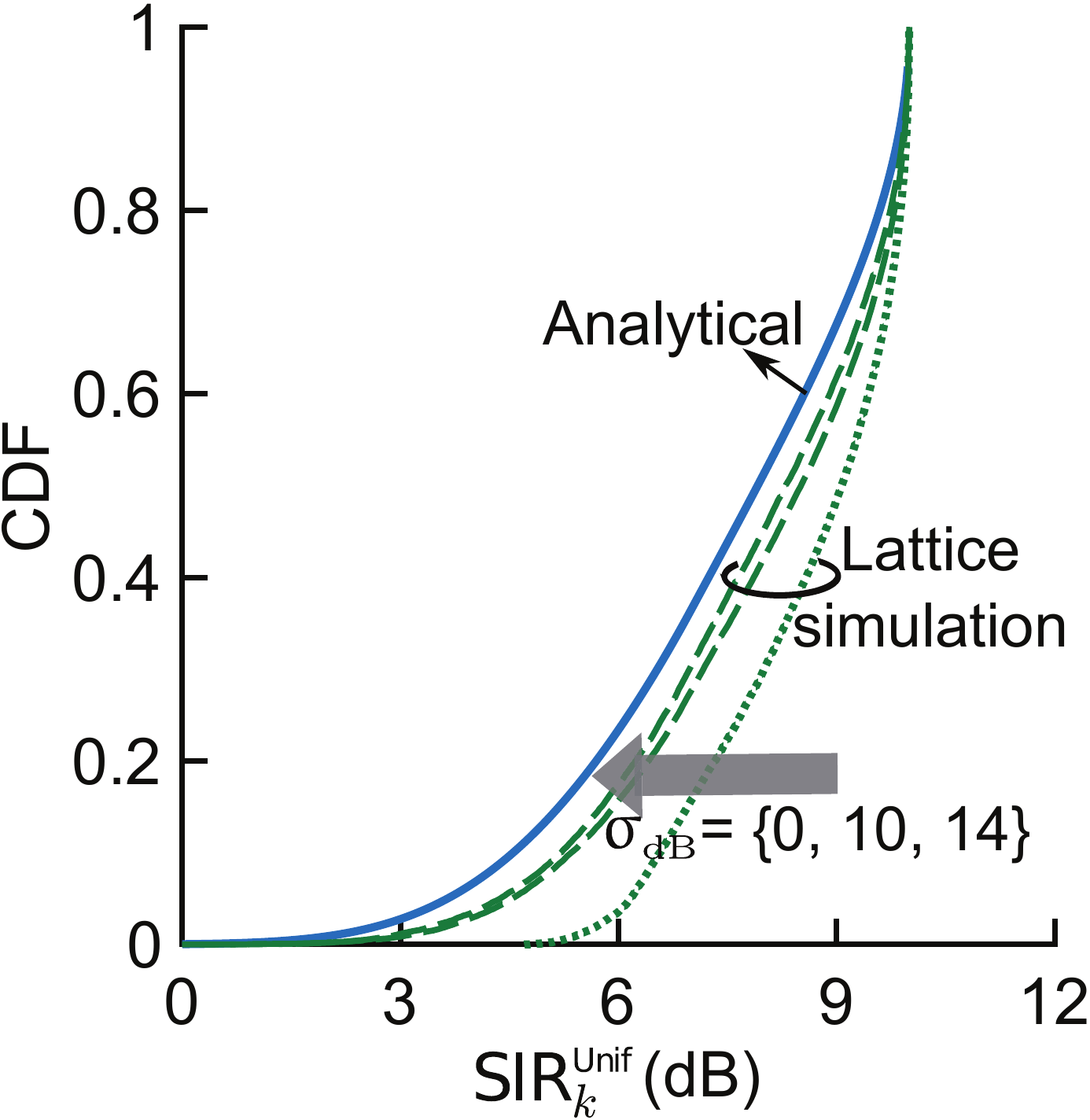}\label{fig:figure2}}
	\quad
	\subfloat[Equal-SIR power allocation; analysis from Prop.~\ref{prp:SIREq_fixedK}.]{\includegraphics[width=0.43\linewidth]{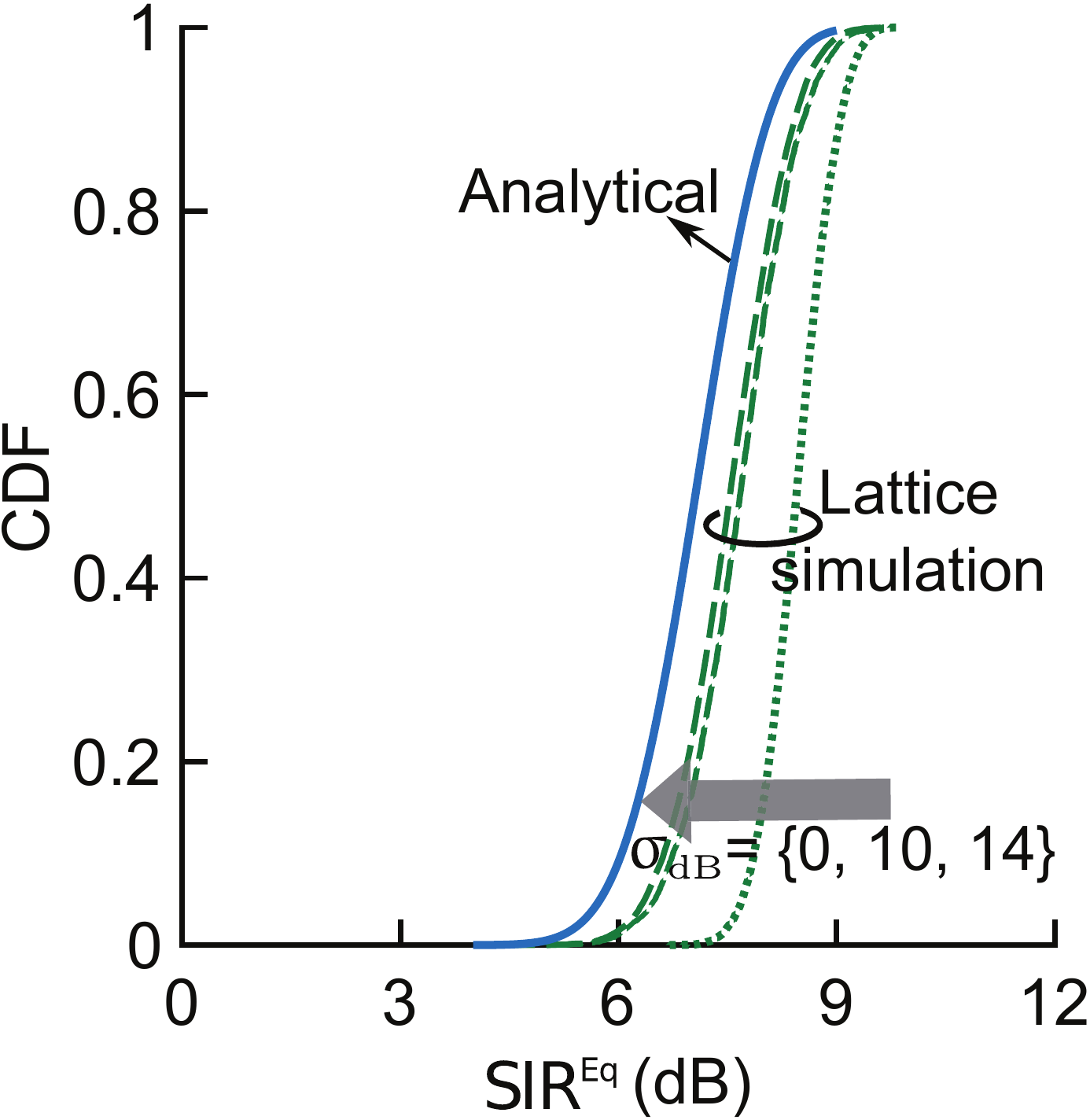}\label{fig:figure3}}
	\caption{CDF of SIR with fixed $K = 10$, $\Na = 100$ and $\eta = 4$. The analytical results, in solid, are contrasted against hexagonal network simulations without ($\shadSD = 0$ dB) and with ($\shadSD = 10$ dB and $14$ dB) shadowing.}
	\label{fig:LatticeNet}
\end{figure}
Let $\Na=100$, $K = 10$ and $\eta =4$. To ensure a fixed $K$, for every network snapshot users are dropped uniformly over the entire network until 
 $K$ are being served by the central BS.
Figs. \ref{fig:figure2}--\ref{fig:figure3} demonstrate the convergence of the SIR CDFs to our characterizations of  $F_{\SIRep_k}(\cdot)$ and $F_{\SIRpa}(\cdot)$ in Props. \ref{Honeymoon1}--\ref{prp:SIREq_fixedK}, respectively.
The dotted and dashed curves are respectively without ($\shadSD = 0$  dB) and with ($\shadSD = 10$ dB and $14$ dB) shadowing in the hexagonal network, while the solid curves correspond to our analytical characterizations.
\end{exmp}


\begin{exmp}\label{ex10}
%
\begin{figure}[!tbp]
	\centering
	\subfloat[Uniform power allocation; analysis from Prop. \ref{prp:SIRUnif_randK}.]{\includegraphics[width=0.43\linewidth]{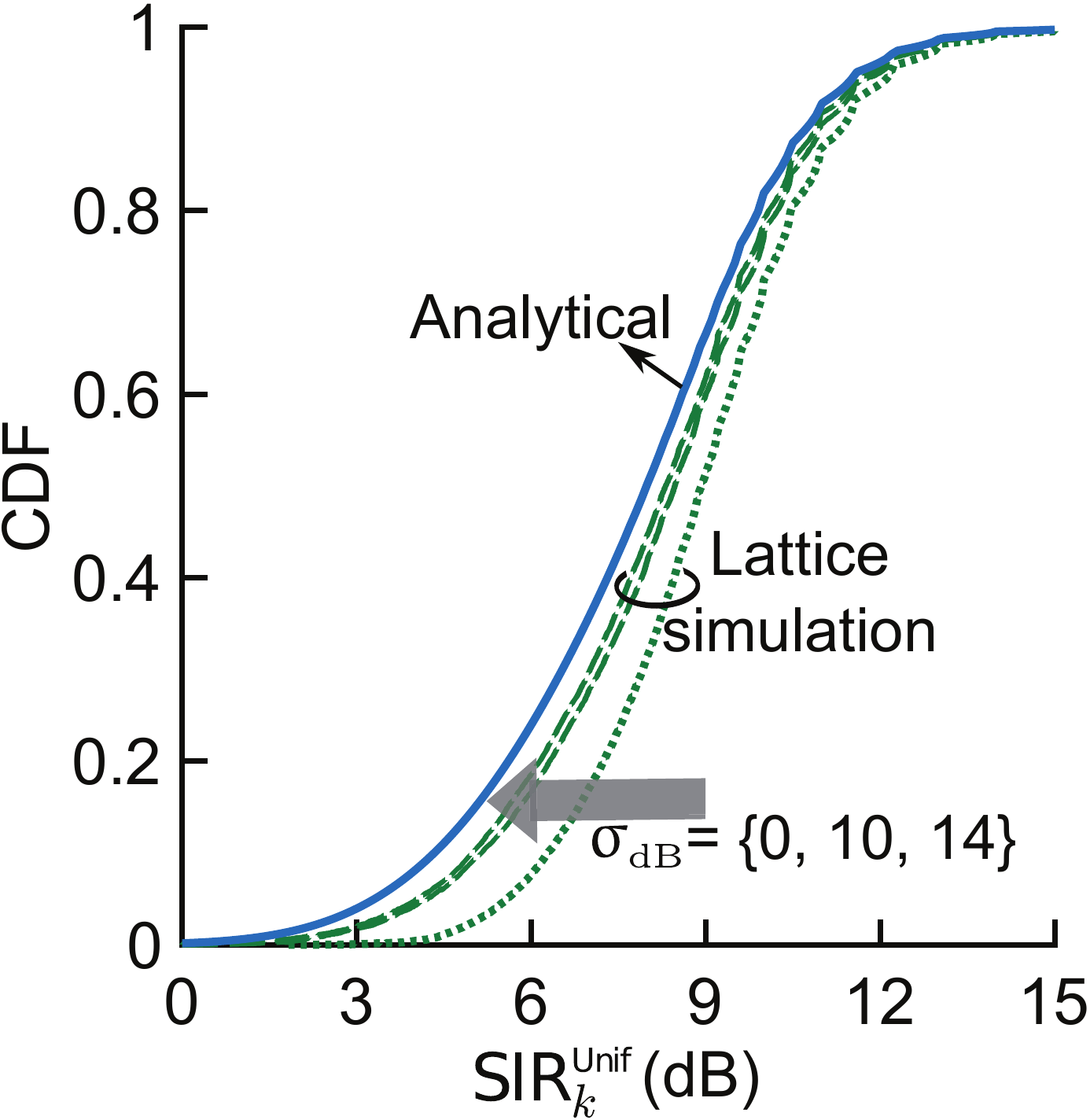}\label{fig:figure7}}
	\quad
	\subfloat[Equal-SIR power allocation; analysis from Prop.~\ref{prp:SIREq_randK}.]{\includegraphics[width=0.43\linewidth]{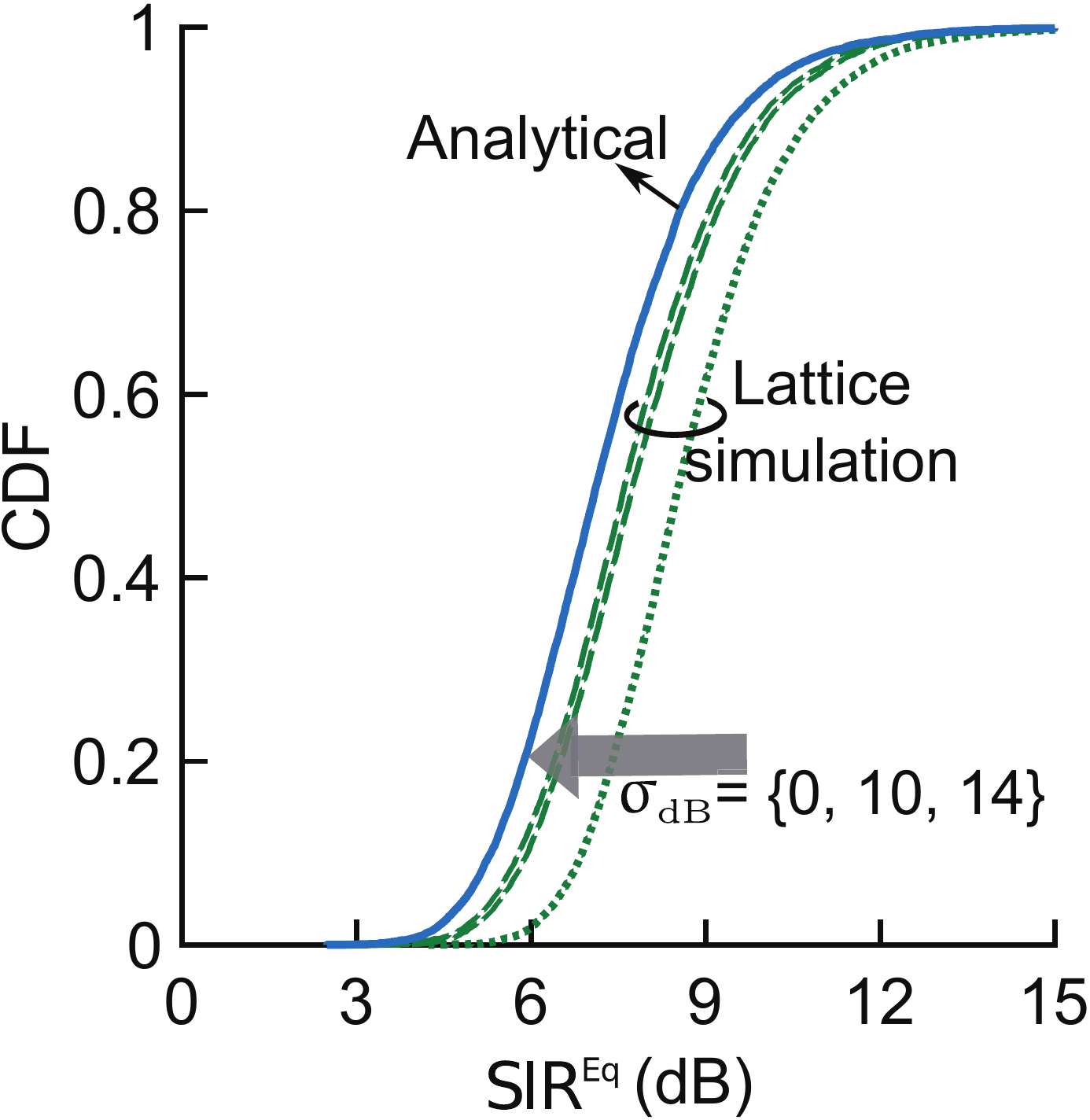}\label{fig:figure8}}
	\caption{CDF of SIR for random $K$, with $\Na = 100$, $\bar{K} = 10$ and $\eta = 4$. The analytical results, in solid, are contrasted against hexagonal network results without ($\shadSD$ = 0 dB) and with ($\shadSD$ = 10 dB and 14 dB) shadowing.}
	\label{fig:LatticeNet2}
\end{figure}
Still on a hexagonal network, let the users now conform to a PPP with density $\bar{K}\lambdab$ while $\Na = 100$, $\bar{K} = 10$ and $\eta = 4$. For each snapshot, the SIRs are computed for the users served by the central BS. (If there are over $100$ such users, a truncation takes place and $100$ of them are selected uniformly at random.)
In Figs. \ref{fig:figure7}--\ref{fig:figure8} we illustrate how, as the shadowing strengthens, the SIR CDFs obtained via simulation in this hexagonal network tend to the corresponding characterizations of  $F_{\SIRep_k}(\cdot)$ and $F_{\SIRpa}(\cdot)$ in Props. \ref{prp:SIRUnif_randK}--\ref{prp:SIREq_randK}, respectively.
\end{exmp}

The closeness of our analytical abstractions (which correspond to $\shadSD  \rightarrow \infty$) to the behaviors with typical outdoor values of $\shadSD$ is conspicuous, similar to the corresponding observation made in the context of nonmassive single-user communication \cite{Ext_TWC}. 
For $\shadSD = 10$ dB, with either power allocation and regardless of whether $K$ is fixed or a truncated Poisson random quantity, the performance in a hexagonal network is within $1$ dB of our analytical characterizations. Furthermore, recalling that $F_{\SIRep_k}(\cdot)$ 
depends on $\Na$ and $K$ only via $\Na/K$, the examples for a uniform power allocation correspond verbatim to any $\Na$ and $K$ related by a factor of $10$.

We further note that Example \ref{ex10} confirms the negligible loss in accuracy incurred by conducting the analysis with a pure Poisson distribution for $K$, rather than the actual truncated distribution.

\subsection{PPP Networks}

For an irregular deployment of BSs and the ensuing variability in cell sizes, let us consider a network where the BS locations are themselves Poisson distributed. Specifically, we consider $\Phib = \Phi \cup \{o\}$, where  $\Phi \subset \mathbb{R}^2$ is a homogeneous PPP 
and $o$ denotes the origin \cite{8168355}. Then, by Slivnyak's theorem, the central BS becomes the typical BS under expectation over $\Phib$. In the Monte-Carlo examples that follow, we drop BSs (500 on average) around the central BS and, as in the previous section, our focus is on the users associated with this central BS. The shadowing is lognormal.

Before proceeding, a relevant observation made in the absence of shadowing (whereby the users are confined to the serving BS's Voronoi cell) is worth reproducing \cite{8168355}: the BS-to-user link distances $\{r_{\ell,(k)}\}$ as perceived by user $k$ in the typical cell are distributed significantly differently from those seen from the typical location on $\mathbb{R}^2$, which underpin $F_\rho(\cdot)$ in our analysis. This effect manifests in the following examples, although only as a very slight deviation from the SIR CDF characterizations, and the convergence with shadowing is rather immediate.

\begin{exmp}

\begin{figure}[!tbp]
		\centering
		\subfloat[Uniform power allocation; analysis from Prop. \ref{Honeymoon1}.]{\includegraphics[width=0.43\linewidth]{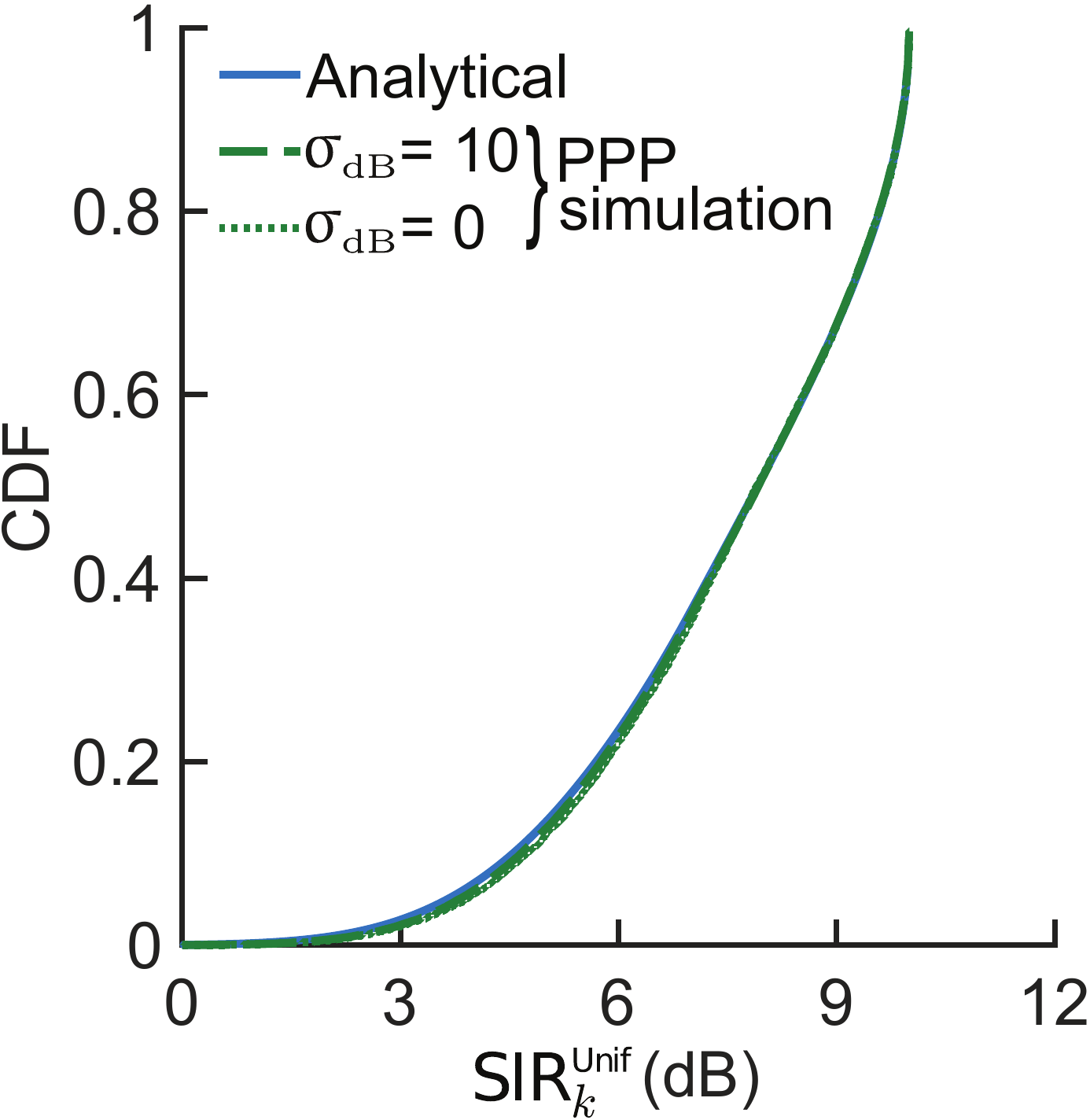}\label{fig:figure9}}
		\quad
		\subfloat[Equal-SIR power allocation; analysis from Prop.~\ref{prp:SIREq_fixedK}.]{\includegraphics[width=0.43\linewidth]{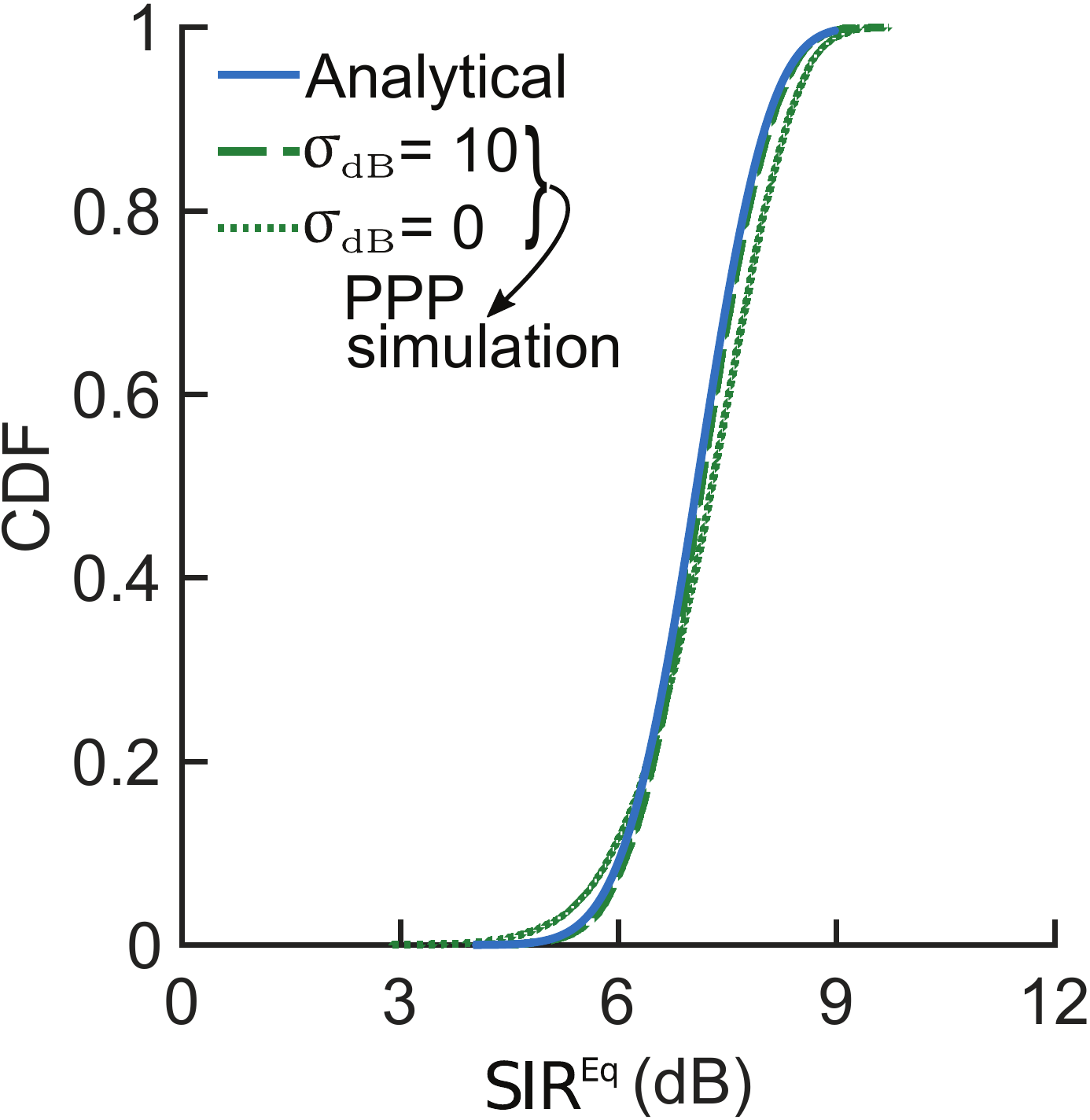}\label{fig:figure10}}
		\caption{CDF of SIR with $K = 10$, $\Na = 100$, and $\eta = 4$. The analytical results, in solid, are contrasted against Poisson network simulation results without ($\shadSD$ = 0 dB) and with ($\shadSD$ = 10 dB) shadowing.}
		\label{fig:PoissonNet}
\end{figure}
With a fixed $K$, again effected for each snapshot of the PPP BSs by uniformly dropping users until $K$ of them are served by the central BS, and with $\Na=100$, $K = 10$ and $\eta =4$, Figs. \ref{fig:figure9}--\ref{fig:figure10} show the convergence of the Monte-Carlo CDFs to our characterizations in Props. \ref{Honeymoon1}--\ref{prp:SIREq_fixedK}, respectively for $F_{\SIRep_k}(\cdot)$ and $F_{\SIRpa}(\cdot)$.
\end{exmp}


In the next example, where $K$ is allowed to be random, we validate the applicability of our results for a Poisson-distributed $K$ (Props. \ref{prp:SIRUnif_randK}--\ref{prp:SIREq_randK}). As advanced in Section \ref{rand num users}, and expounded in more detail in \cite{GLH18}, the simulated $K$ acquires its Poisson nature only as the shadowing intensifies.
This causes a gap between analysis and simulation in the absence of shadowing, but this gap closes quickly as shadowing makes its appearance (cf. Figs. \ref{joppu}--\ref{freddy}).


\begin{exmp}\label{ex12}
	
	\begin{figure}[!tbp]
		\centering
		\subfloat[Uniform power allocation; analysis from Prop. \ref{prp:SIRUnif_randK}.]{\includegraphics[width=0.43\linewidth]{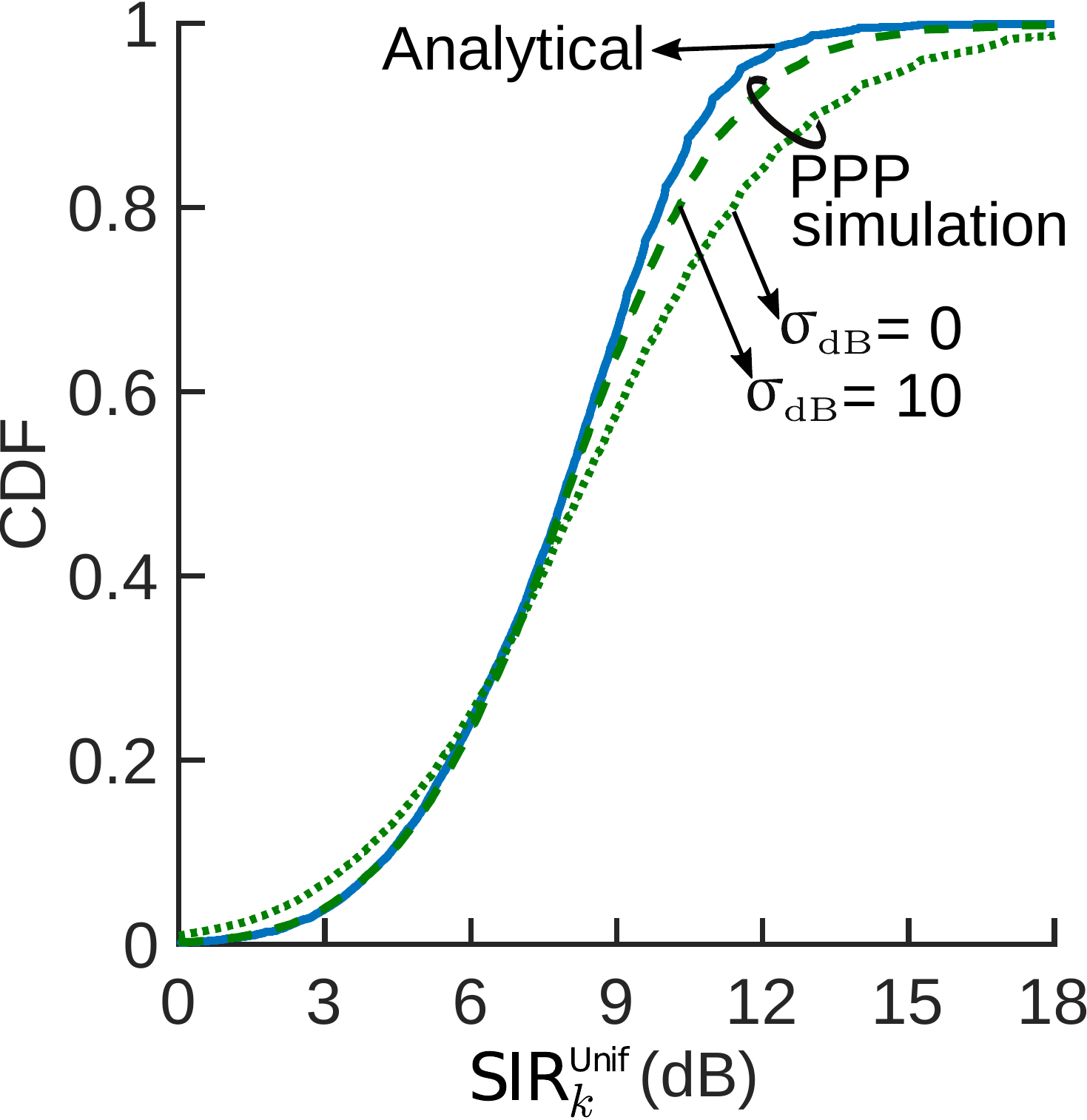}\label{joppu}}
		\quad
		\subfloat[Equal-SIR power allocation; analysis from Prop.~\ref{prp:SIREq_randK}.]{\includegraphics[width=0.43\linewidth]{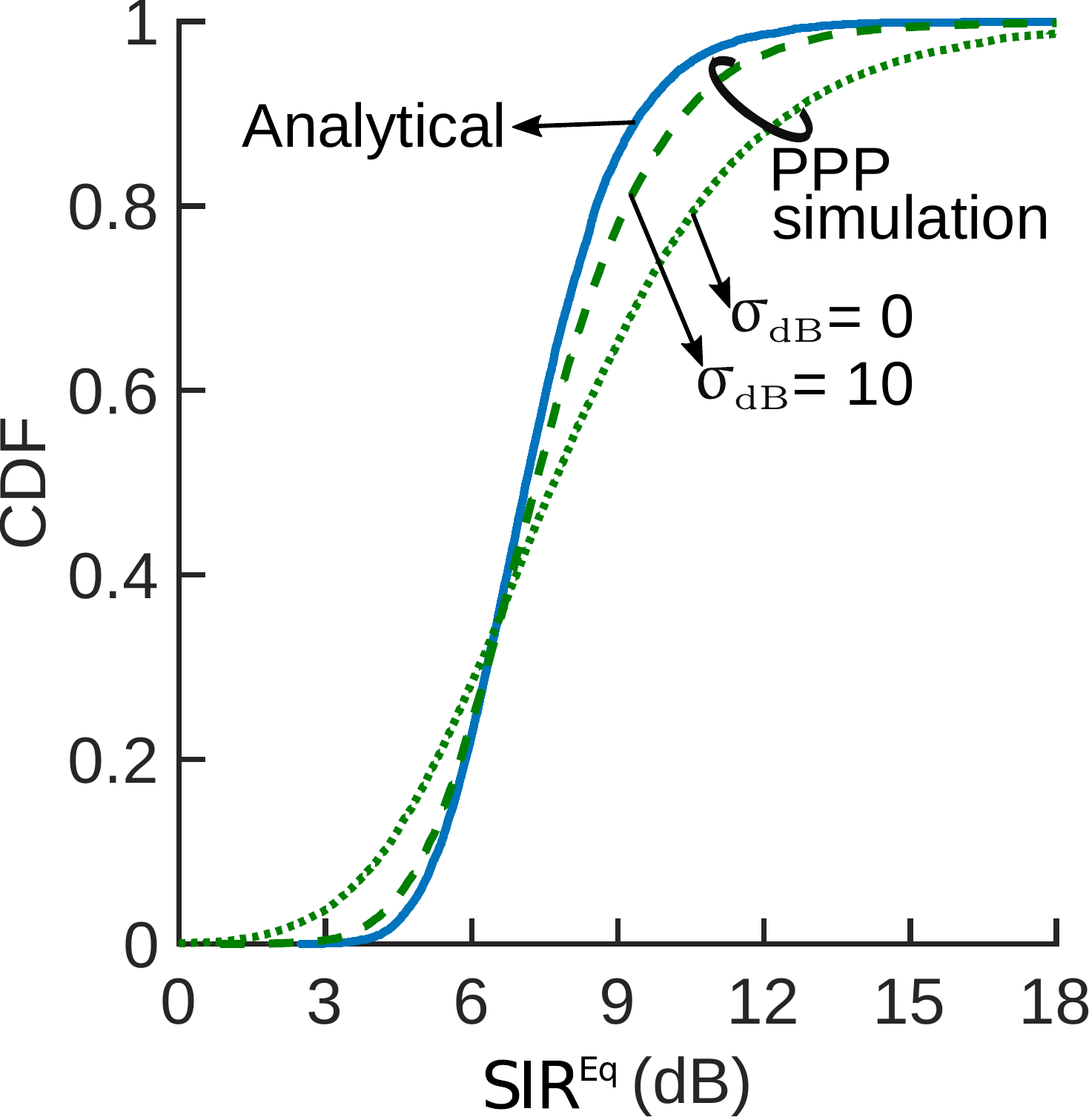}\label{freddy}}
		\caption{CDF of SIR under random $K$, with $\Na = 100$, $\bar{K} = 10$, and $\eta = 4$. The analytical results, in solid, are contrasted against PPP network simulation results without ($\shadSD$ = 0 dB) and with ($\shadSD$ = 10 dB) shadowing.}
		\label{fig:PoissonNet2}
	\end{figure}
	
	Consider again a PPP network of BSs, with the users conforming also to a PPP of density $\bar{K}\lambdab$, and with $\Na = 100$, $\bar{K} = 10$ and $\eta = 4$. 
	For each snapshot, the PPPs of BSs and users are realized over the whole network region
	and the SIRs are computed for the users served by the central BS with their number again truncated at $K= 100$.
	In Figs. \ref{joppu}--\ref{freddy} we show how, as the shadowing strengthens, the SIR CDFs obtained via Monte-Carlo quickly tend to the corresponding characterizations in Props. \ref{prp:SIRUnif_randK}--\ref{prp:SIREq_randK}, respectively for $F_{\SIRep_k}(\cdot)$ and $F_{\SIRpa}(\cdot)$.
\end{exmp}

\section{Impact of Noise and Pilot Contamination}
\label{tooth}

\subsection{Noise}
\label{AWGN}

The interference-limited regime reflects well the operating conditions of mature systems, justifying our SIR-based analysis with thermal noise neglected.
Nevertheless, for the sake of completeness, we herein corroborate that noise does not significantly alter the applicability of our expressions.

The signal-to-interference-plus-noise ratio (SINR) at the $k$th user can be formulated as \cite[Sec. 10.5.1]{Heath-Lozano-18}
\begin{align}
\SINR_k  &= \frac{\Na \frac{P_k}{P} \, G_{(k)}^2}{\left(G_{(k)} + \snrratio \, \noiseVar/P \right)\left(\sum_{\ell}G_{\ell,(k)} +\noiseVar/P \right)},
\end{align}
where $\snrratio$ is the ratio between the forward- and reverse-link signal-to-noise ratios and
the reverse-link pilots are assumed not to be power-controlled.
With a uniform power allocation, the SINR specializes to 
\begin{align}\label{eq:UnifnoiseSIR}
\SINRep_k 
&= \frac{\Na/K}{\left(1 + \snrratio \frac{\noiseVar}{P \, G_{(k)}}\right)\left(1 +  1/\rho_k + \frac{\noiseVar}{P \, G_{(k)}}\right)}, 
\end{align}
where, recall, $\rho_k = G_{(k)}/\sum_{\ell\neq0} G_{\ell,(k)}$.
In turn, with an equal-SINR power allocation,
\begin{align}\label{eq:EqnoiseSIR}
\SINRpa &= \frac{\Na}{\sum_{k=0}^{K-1}\left(1 + \snrratio \frac{\noiseVar}{P \, G_{(k)}}\right)\left(1 +  1/\rho_k + \frac{\noiseVar}{P \, G_{(k)}}\right)}.
\end{align}

\begin{exmp}
\begin{figure}[!tbp]
	\centering
	\subfloat[Uniform power allocation; analysis from Prop. \ref{prp:SIRUnif_randK}.]{\includegraphics[width=0.43\linewidth]{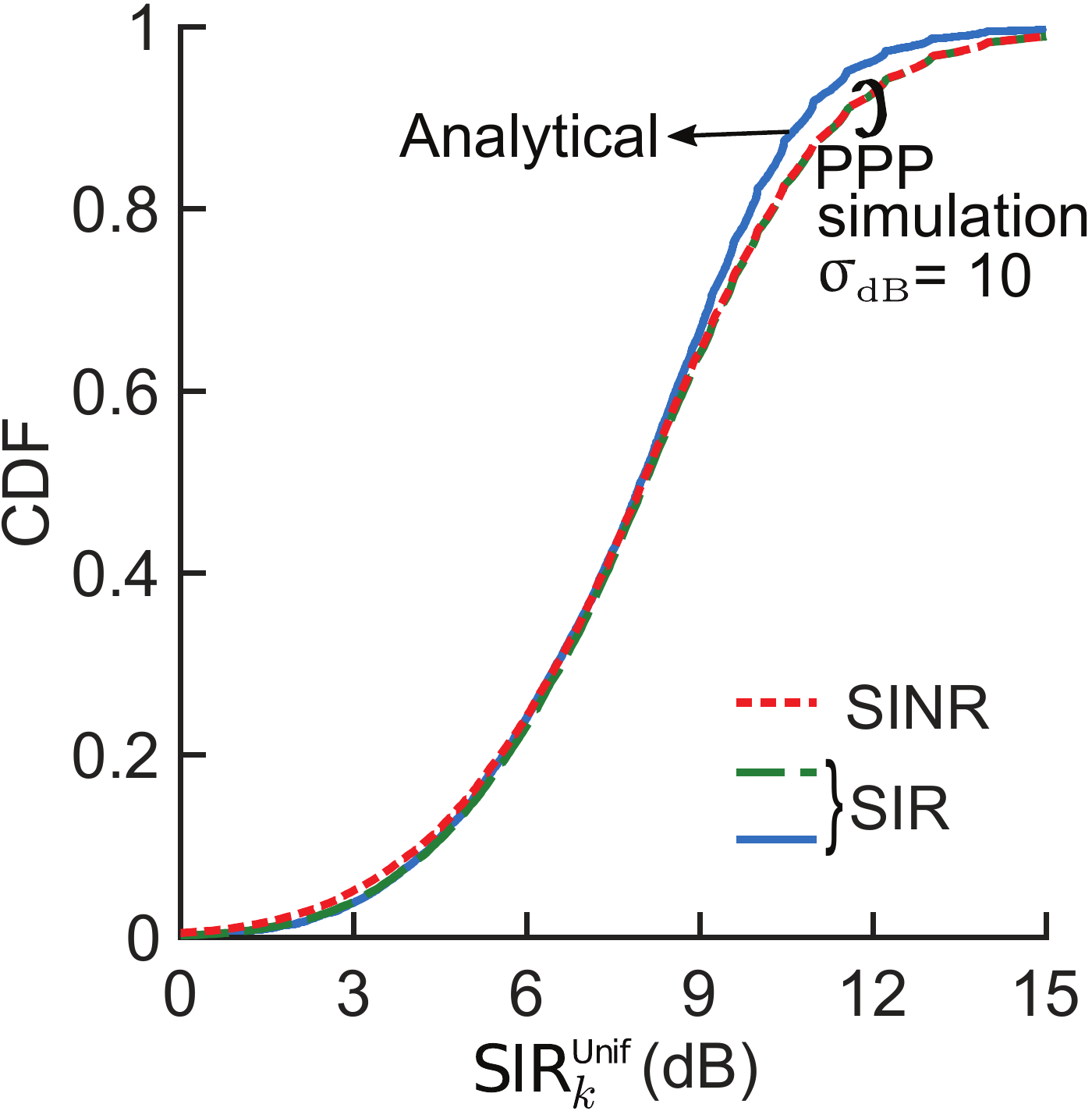}\label{fig:noiseSIRUnif}}
	\quad
	\subfloat[Equal-SIR power allocation; analysis from Prop.~\ref{prp:SIREq_randK}.]{\includegraphics[width=0.43\linewidth]{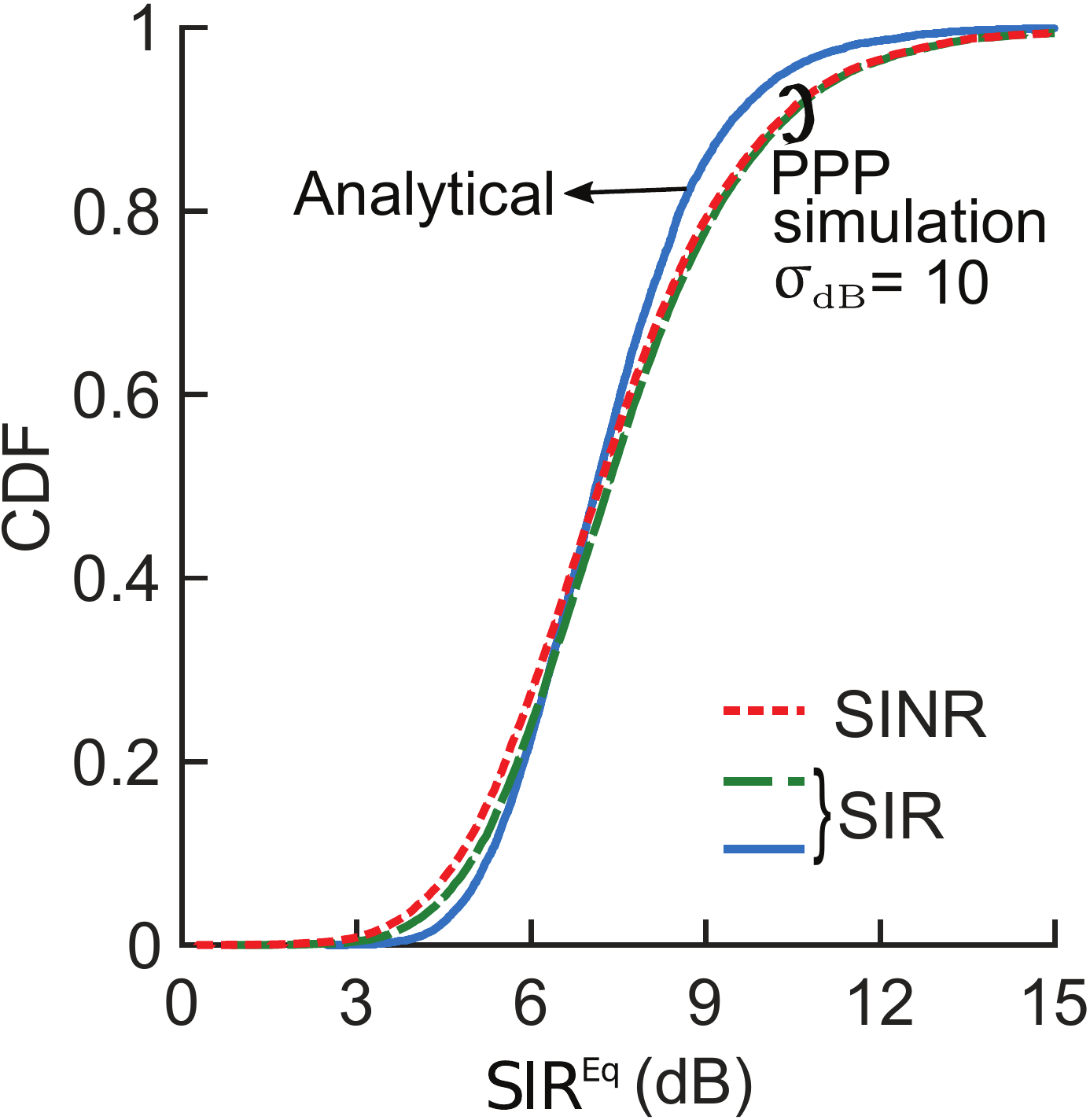}\label{fig:noiseSIREq}}
	\caption{CDFs of SINR and SIR under random $K$, with $\Na = 100$, $\bar{K} = 10$ and $\eta = 4$. The simulated SINR is contrasted against the SIR, both analytical and simulated.}
\label{fig:noiseSIR}
\end{figure}
Reconsider Example \ref{ex12} with
$\lambdab=1$ BS$/\text{km}^2$. The path loss intercept is $\Lref = -128$ dB at $1$ km, which is a reasonable value for such BS density at around $2$ GHz.
Each BS emits $P = 47.8$ dBm on the forward link, the reverse-link pilot power is $20$ dB lower, and the noise figures at the user and BS receivers are $7$ and $3$ dB, respectively, such that $\snrratio = 16$ dB.
The bandwidth is the standard $20$ MHz of LTE, the noise power spectral density is $N_0 = -174$ dBm/Hz, and the BS antennas feature $6$ dBi of vertical gain, which adds to $G_{(k)}$. The CDFs of $\SINRep_k$ and $\SINRpa$, respectively simulated via (\ref{eq:UnifnoiseSIR}) and (\ref{eq:EqnoiseSIR}) with shadowing ($\shadSD = 10$), are plotted in Fig. \ref{fig:noiseSIR} alongside their SIR counterparts replotted from Fig. \ref{fig:PoissonNet2}.
\end{exmp}

As in the foregoing example, an excellent match between the SINRs and SIRs is observed for all other cases reported in the paper. 

\subsection{Pilot Contamination}
\label{contamination}

Finally, in order to gauge the deviations that pilot contamination may bring about,
we contrast our solutions with
simulations that do incorporate contamination. 
Denoting by $\mathcal{P}$ the set of indices of the BSs reusing the pilots of the $k$th user of interest, 
when the LMMSE (linear minimum mean-square error) channel estimation incorporates the ensuing contamination \cite[Sec. 10.5.1]{Heath-Lozano-18}
\begin{align}
\SIR_k  &= \frac{\frac{\Na}{ G_{(k)} + \sum_{\ell \in \mathcal{P}} G_{(\ell,k)}} \frac{P_k}{P} \, G^2_{(k)}}{\sum_\ell G_{\ell,(k)} + \sum_{\ell \in \mathcal{P}} \frac{\Na }{G_{\ell,(k)} + \sum_{l \in \mathcal{P}} G_{\ell,(l,k)}} \frac{P_{\ell,k}}{P} \, G^2_{\ell,(k)} },
\end{align}
where the reverse-link pilots are again assumed not to be power-controlled.
%
%
With a uniform power allocation, and with a fixed $K$ in all the contaminating cells, the above expression specializes to
\begin{align}
\SIRep_k  
&= \frac{\frac{\Na/K}{1 + \sum_{\ell \in \mathcal{P}}  G_{(\ell,k)}/G_{(k)}}}{ 1 + 1/\rho_k + \sum_{\ell \in \mathcal{P}} \frac{G_{\ell,(k)}}{G_{(k)}} \frac{\Na/K}{1 + \sum_{l \in \mathcal{P}}  G_{\ell,(l,k)}/G_{\ell,(k)}}},  \label{eq:contaminationSIR}
\end{align}
%
%
while, with an equal-SIR power allocation,
\begin{align}\label{eq:EqcontSIR}
\SIRpa  &= \frac{\Na}{\sum_{k=0}^{K-1} \left(1 + \sum_{\ell \in \mathcal{P}}  G_{(\ell,k)}/G_{(k)}\right) \left( 1 + 1/\rho_k + \sum_{\ell \in \mathcal{P}} \frac{G_{\ell,(k)}}{G_{(k)}} \frac{\Na/K}{1 + \sum_{l \in \mathcal{P}}  G_{\ell,(l,k)}/G_{\ell,(k)}} \right)}  .
\end{align}

\subsubsection{Hexagonal Lattice Networks}

\begin{figure}[!tbp]
	\centering
	\subfloat[Pilot reuse factor $\Lc = 4$.]{\includegraphics[width=0.36\linewidth]{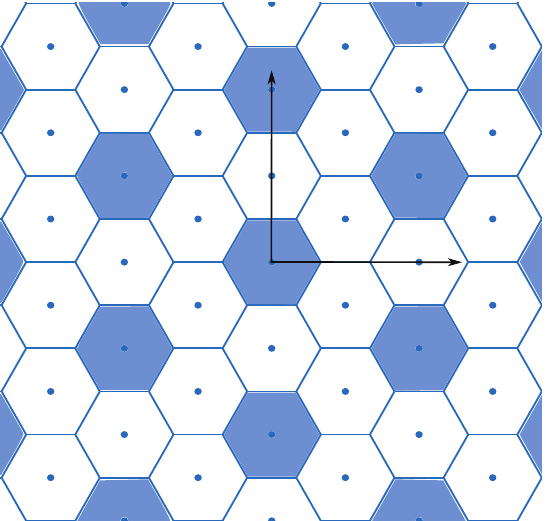}\label{fig:Lc4pat}}
	\quad
	\subfloat[Pilot reuse factor $\Lc = 7$.]{\includegraphics[width=0.36\linewidth]{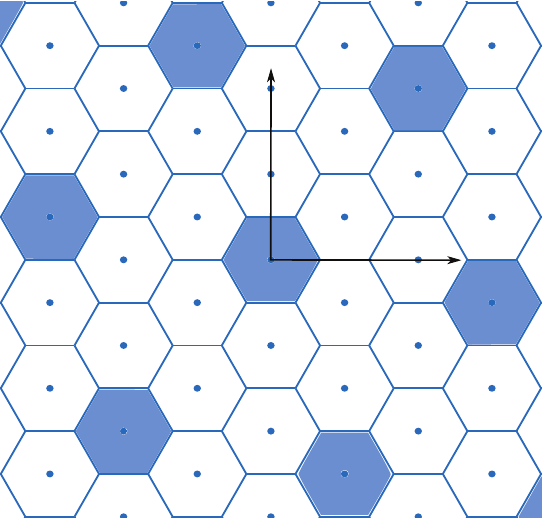}\label{fig:Lc7pat}}
	\caption{Pilot reuse patterns in a hexagonal grid network: the shaded cells reuse the same pilots.}
	\label{fig:RUpat}
\end{figure}

In lattice networks, regular pilot reuse patterns can be established.
Precisely, we consider pilot reuse patterns whereby each pilot is used in only one cell within every cluster of $\Lc$ adjacent cells.
Such reuse patterns, illustrated in Fig.~\ref{fig:RUpat} for a hexagonal network with $\Lc = 4$ and $\Lc = 7$, are implementationally convenient.
And they are suboptimum in the face of shadowing, meaning that they are a conservative choice for our assessment; with a dynamic pilot assignment, the impact of contamination would abate even further.

\begin{exmp}\label{exmp:contamination}
	\begin{figure}[!tbp]
		\centering
		\subfloat[Uniform power allocation; analysis from Prop. \ref{Honeymoon1}.]{\includegraphics[width=0.43\linewidth]{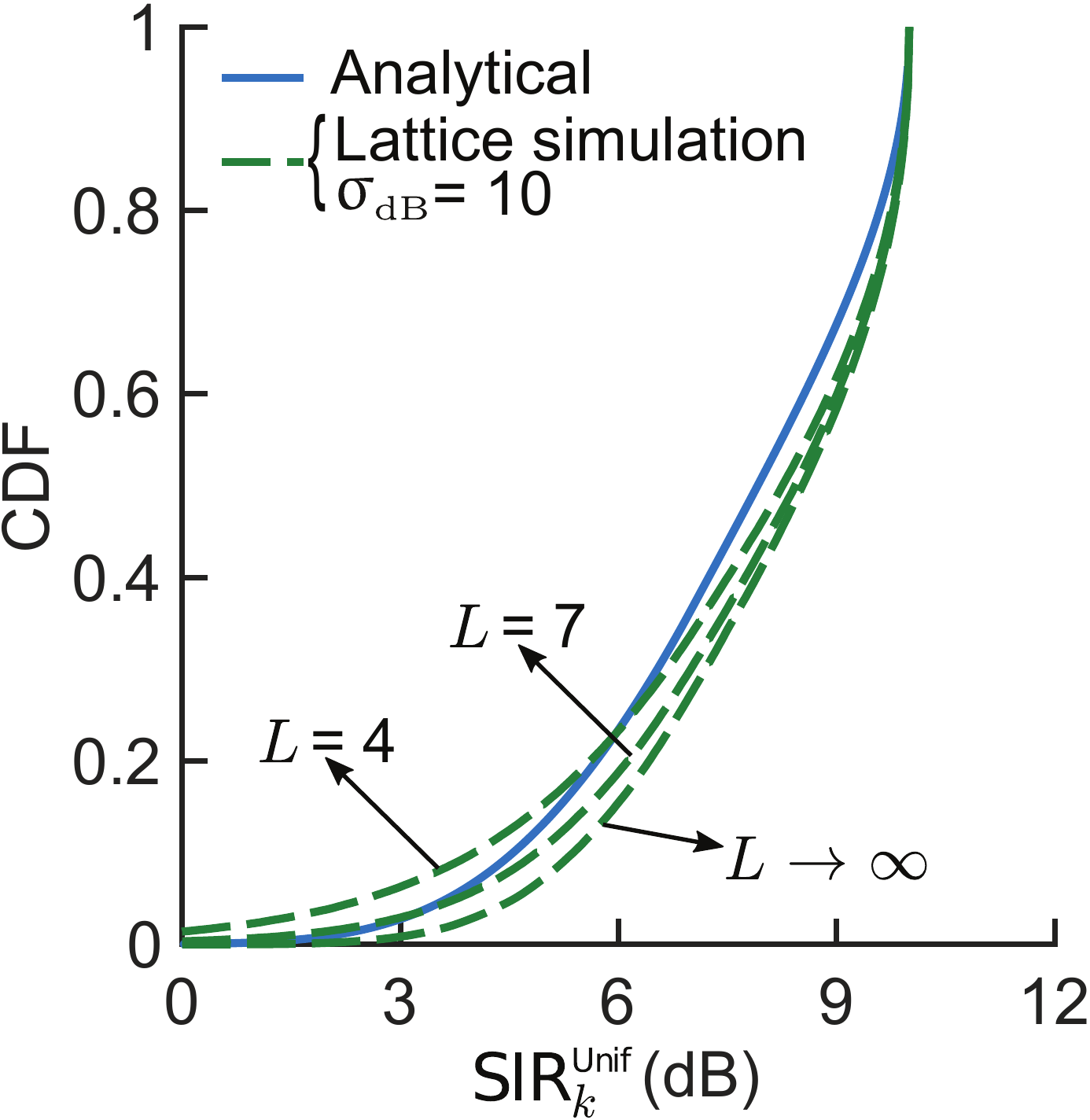}\label{fig:contaminationSIRUnif}}
		\quad
		\subfloat[Equal-SIR power allocation; analysis from Prop.~\ref{prp:SIREq_fixedK}.]{\includegraphics[width=0.43\linewidth]{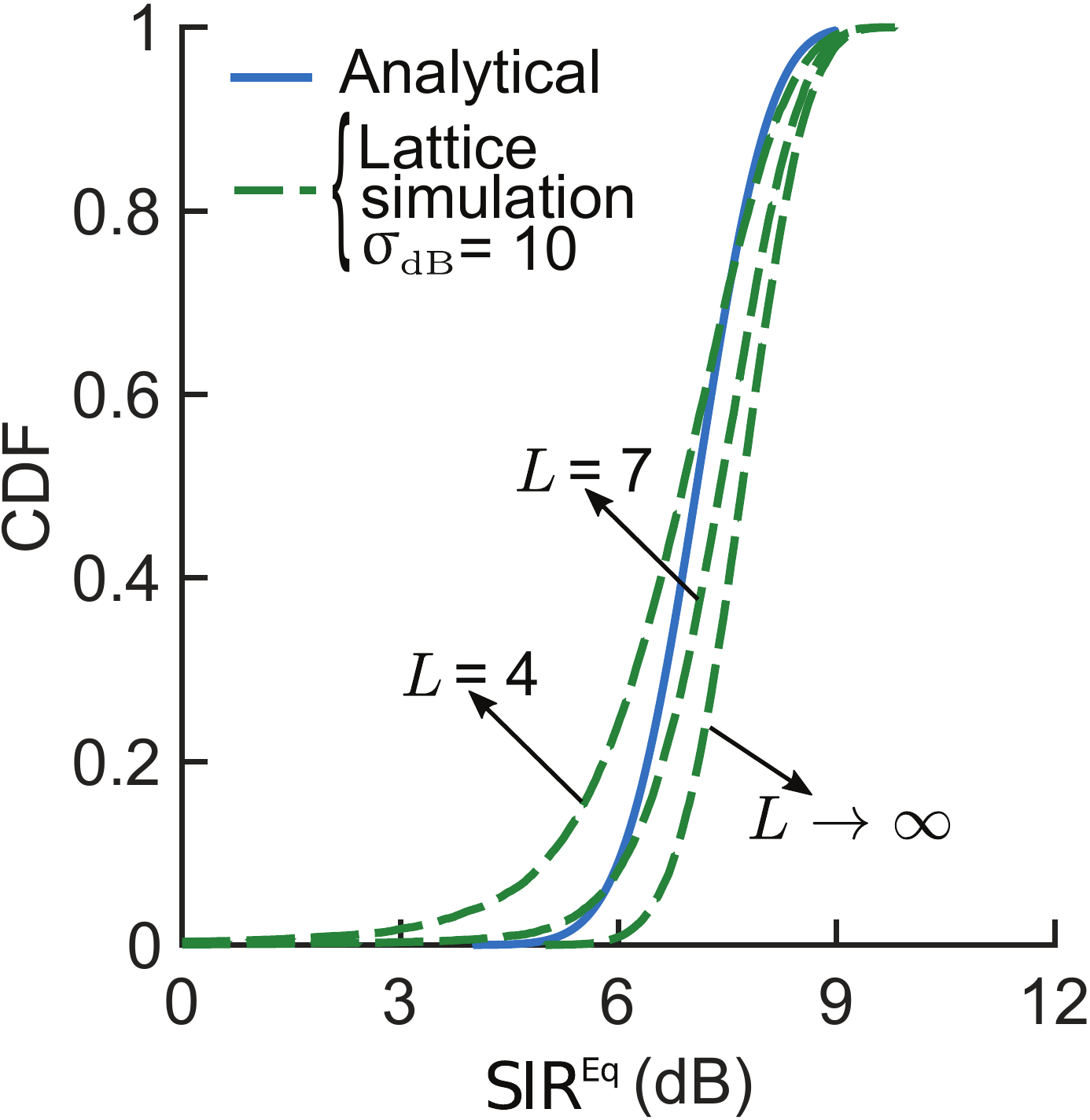}\label{fig:contaminationSIREq}}
		\caption{CDFs of SIR with fixed $K = 10$, $\Na = 100$ and $\eta = 4$. The hexagonal network simulations incorporate pilot contamination via (\ref{eq:contaminationSIR}) and (\ref{eq:EqcontSIR}), respectively for $\SIRep_k$ and $\SIRpa$, in addition to shadowing ($\shadSD$ = 10 dB).}
		\label{fig:contaminationSIR}
	\end{figure}
Consider again $\eta =4$, $\Na = 100$ and $K = 10$. CDFs of $\SIRep_k$ and $\SIRpa$ with pilot contamination accounted for, respectively through (\ref{eq:contaminationSIR}) and (\ref{eq:EqcontSIR}), and computed via Monte-Carlo for a hexagonal network with shadowing ($\shadSD$ = 10 dB) are plotted in Figs. \ref{fig:contaminationSIRUnif}--\ref{fig:contaminationSIREq}. The results for pilot reuse factors $\Lc = 4$ and $\Lc = 7$ are plotted alongside their counterparts that forgo contamination (replotted from Fig. \ref{fig:LatticeNet}), namely the hexagonal simulation result with $\Lc \rightarrow \infty$ and the analytical curve.
\end{exmp}

For $\Lc = 7$, and even for $\Lc=4$, the contamination is seen to have a minor effect.
Our analysis is therefore applicable provided these reuse factors are feasible, and that is indeed the case over a broad operating range.
For a fading coherence of $\Nc=1000$ symbols, which is a reasonably low value,\footnote{This value is obtained as $\Nc = \ceil{\frac{\mathsf{v}}{\uplambda \, B_{\rm c}}}$ with a wavelength of $\lambda = 0.15$ m, a coherence bandwidth of $B_{\rm c} = 370$ kHz, and a velocity of $\mathsf{v} = 100$ km/h.}
the pilot overhead equals $K \Lc / 1000$. For $\Lc=4$ and $\Lc=7$, and with the values of $K$ corresponding to $\Na = 100$ or even $\Na=200$ antennas, such overhead is perfectly acceptable.
For a condition determining the combinations of $\Na$, $K$, and $\eta$ under which the contamination has a minor effect with tighter pilot reuse patterns ($\Lc=3$ or even $\Lc=1$), readers are referred to \cite[Example 10.4]{Heath-Lozano-18}.

\subsubsection{PPP Networks}

In nonlattice networks, the establishment of pilot reuse patterns requires site-specific planning on the part of the operator, or else some clustering algorithm \cite{attarifar2018random}. 
As such schemes are well beyond the scope of our paper, the next example abstracts them and gauges the impact of contamination in a manner that is not specific to any pilot allocation approach.


\begin{exmp}
\begin{figure} 
		\centering
		\includegraphics [width = 0.65\columnwidth]{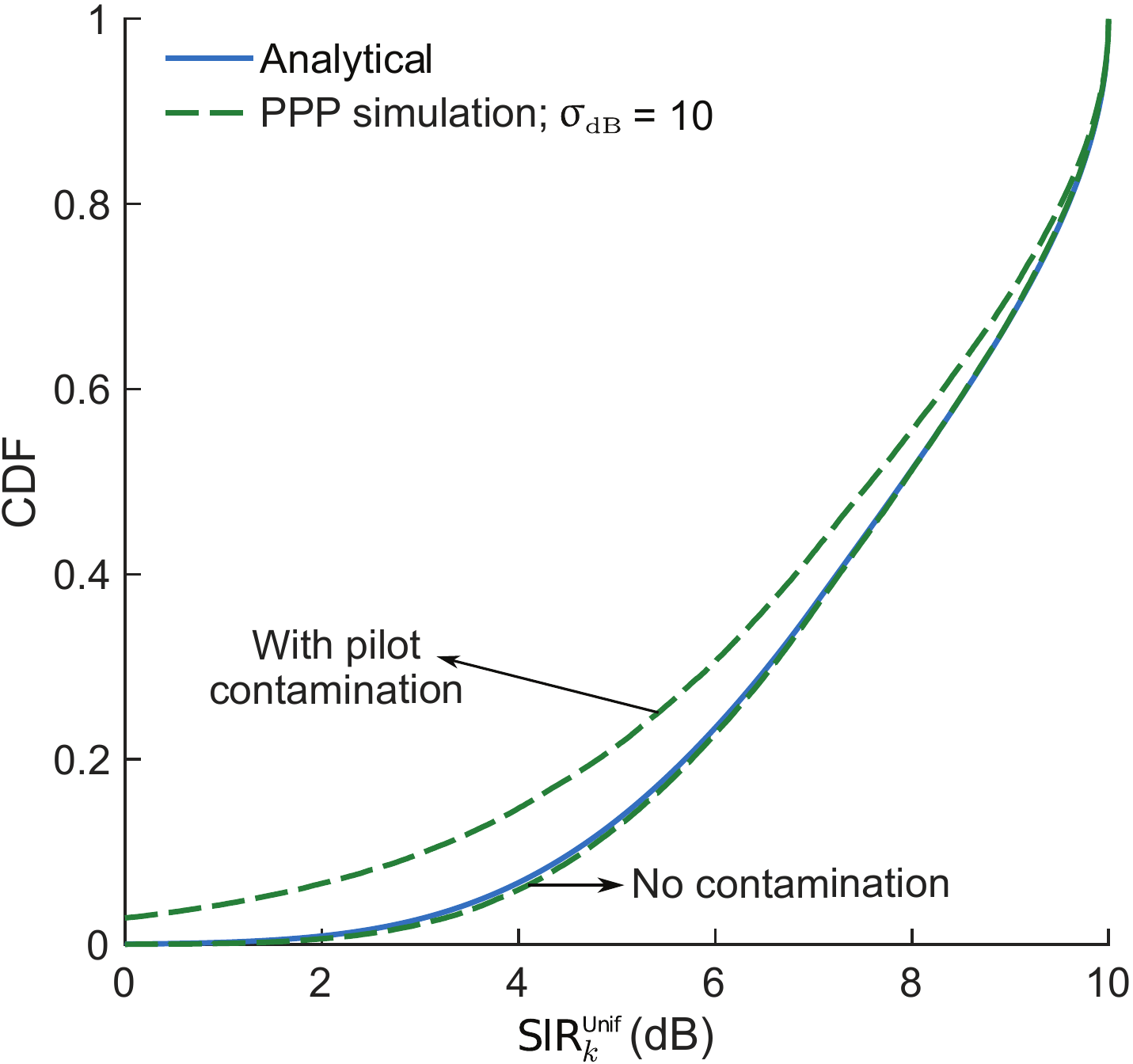} 
		\caption{CDF of $\SIRep_k$ for $\Na/K = 10$ and $\eta = 4$ in a PPP network. The contamination is incorporated to the simulations via (\ref{eq:contaminationSIR}) with an average reuse factor of $\Lc=8.5$. The analytical curve is from Prop. \ref{Honeymoon1}.}
		\label{fig:contaminationPPP}
\end{figure}
Consider again a PPP network with $\eta = 4$, $\Na = 100$, $K = 10$ and $\shadSD = 10$.
For each network realization, out of 500 random pilot allocations (wherein each interfering BS belongs to $\mathcal{P}$ with probability 1/7), we select the one minimizing the average contamination power per BS, computed as
\begin{align}
\frac{1}{|\mathcal{P}_0|}\sum_{l \in \mathcal{P}_0} \sum_{\ell \in \mathcal{P}_0, \ell \neq l} \sum_{k=0}^{K-1} G_{\ell,(l,k)},
\end{align}
where $\mathcal{P}_0 = \{0\} \cup \mathcal{P}$ is the set of all mutually contaminating BSs including the central BS while $|\mathcal{P}_0|$ is its cardinality.
The effective pilot reuse factor becomes $\Lc=8.5$ on average.

In Fig. \ref{fig:contaminationPPP}, the CDF of $\SIRep_k$ with pilot contamination, simulated via (\ref{eq:contaminationSIR}) with the aforedescribed procedure, is plotted along side its contamination-free counterpart (replotted from Fig. \ref{fig:figure9}) and the analytical curve from Prop. \ref{Honeymoon1}.

\end{exmp}

\section{Conclusion}
\label{summary}

While formally derived for interference-limited networks devoid of pilot contamination and asymptotic in the shadowing strength, the expressions presented in this paper for conjugate beamforming are broadly applicable, as confirmed by a multitude of examples. This broad applicability of the analysis
makes its extension to other transmit and receive strategies, power allocations, user association policies, and even channel conditions, highly desirable.




\section*{Acknowledgment}
The efficient editorial handling by Prof. Kaibin Huang and the excellent feedback provided by the reviewers are gratefully acknowledged.


\appendices
\section{Proof of Lemma \ref{lem:charfunrho}}
\label{app:proof of lemma1}

Capitalizing on the fact that the characteristic function, $
\varphi_X(\omega) = \mathbb{E}[e^{i \omega X}]$, can be inverted as per the Gil-Pelaez theorem \cite{Gil-Pelaez} to compute
\begin{align}\label{joph}
\mathbb{P}[X<\mathsf{x}] = \frac{1}{2} - \frac{1}{\pi} \int_{0}^{\infty} \Im \{e^{-i\mathsf{x \omega}} \varphi_X(\omega)\} \frac{d\omega}{\omega},
\end{align}
we write
\begin{align}
F_\rho(\theta) &= \mathbP \big[1 - \theta/\rho < 0 \big] \\
&= \frac{1}{2} - \frac{1}{\pi} \int_{0}^{\infty} \Im \big \{\varphi_{1 -\theta/\rho}(\omega)\big \} \frac{d\omega}{\omega} \\
&= \frac{1}{2} - \frac{1}{\pi} \int_{0}^{\infty} \Im\big \{ e^{i\omega} \varphi_{1/\rho}(-\theta\,\omega)\big\} \frac{d\omega}{\omega} \\
&= \frac{1}{2} - \frac{1}{\pi} \int_{0}^{\infty} \Im \! \left\{e^{i\omega} \mathbE \! \left[e^{-i\theta\omega/\rho}\right] \right\} \frac{d\omega}{\omega} \\
&= \frac{1}{2} - \frac{1}{\pi} \int_{0}^{\infty} \Im \! \left\{e^{i\omega} \mathcal{L}_{1/\rho}(i\theta\omega) \right\} \frac{d\omega}{\omega} \label{baba}
\end{align}
and plug the Laplace domain characterization \cite[Lemma 2]{ICC16-Ganti-Haenggi}
\begin{align}
\mathcal{L}_{1/\rho}(\mathsf{s}) &= \mathbb{E}\left[e^{-\mathsf{s}/\rho}\right] \\
&= \frac{1}{{}_1 F_1(-\delta,1-\delta,-\mathsf{s})} \\
&= \frac{e^\mathsf{s}}{{}_1 F_1(1,1-\delta,\mathsf{s})}  \label{eq:LTrho0}
\end{align}
into (\ref{baba}) to complete the proof.

\section{CDF via Numerical Inversion of Laplace Transform}\label{app:NumLapInv}
	\begin{lem} \cite[Prop. 1]{7880695}
		\label{lem:ch2_Frho}
		The CDF of $\rho$ can be computed as
		\begin{align}
		\label{eq:rhoCDF_laplaceiv}
		F_{\rho}(\theta) \approx 1 - \frac{e^{A/2} }{2^\mathsf{B}} \, \sum_{b=0}^{\mathsf{B}} \binom{\mathsf{B}}{b} \sum_{q=0}^{\mathsf{Q}+b} \frac{ (-1)^q }{D_q} \, \Re \! \left\{\frac{\mathcL_{1/\rho}(\theta \, t) }{t} \right\}
		\end{align}
		where $t = ( A + \imunit 2 \pi q)/2$ while $D_0 = 2$ and $D_q=1$ for $q \geq 1$, and recall that $\mathcal{L}_{1/\rho}(\cdot)$ is in (\ref{eq:LTrho0}). In turn, the parameters $A$, $\mathsf{B}$ and $\mathsf{Q}$ control the accuracy of the approximation.
		Suggested values for numerical Laplace transform inversions with many digits of precision are $A=18.4$, $\mathsf{B}=11$ and $\mathsf{Q}=15$ \cite{Numerical-Laplace-inversion}. Following the recommendations of \cite{Euler-Laplace-inversion} for a 4-digit accuracy, a more relaxed---but still plentiful for our purposes---precision target, we obtain $A=9.21$ and $\mathsf{B}=5$, with $\mathsf{Q}$ as large as possible; as it turns out, moderate values of $\mathsf{Q}$ suffice to yield error levels that are negligible.
		\end{lem}
		
\section{Proof of Proposition \ref{prp:SIREq_fixedK}}\label{app:SIREq_fixedK}

Invoking (\ref{joph}),
\begin{align}
F_{\SIRpa}(\theta)
&= \mathbP\left[1 -\theta/\SIRpa < 0\right] \\
&= \frac{1}{2} - \frac{1}{\pi} \int_{0}^{\infty} \Im \! \left\{\varphi_{1 -\theta/\SIRpa}(\omega)\right\} \frac{d\omega}{\omega} \\
&= \frac{1}{2} - \frac{1}{\pi} \int_{0}^{\infty} \Im \! \left\{e^{i\,\omega} \, \varphi_{1/\SIRpa}(-\theta\omega)\right\} \frac{d\omega}{\omega},
\end{align}
where the characteristic function $\varphi_{1/\SIRpa} (\omega) = \mathbb{E}  [e^{i\omega/\SIRpa}]$ can be computed as
\begin{align}
\varphi_{1/\SIRpa} (\omega) 
&= 	\mathbb{E} \! \left[e^{\frac{i \omega}{\Na} \left(K + \sum_{k=0}^{K-1}1/\rho_k\right)} \right]  \\
&= e^{\frac{i \omega  K}{\Na}} \,  \mathbb{E}\!\left[\prod_{k=0}^{K-1} e^{\frac{i  \omega}{\Na \, \rho_k}}\right]  \\
&= e^{\frac{i  \omega  K}{\Na}} \left(\mathbb{E}_{\rho}\!\left[e^{\frac{i  \omega}{\Na \, \rho}}\right]\right)^K \label{eq:step1}\\
&= {}_1 F_1\!\left(1,1-\delta,-i \,\omega/\Na\right)^{-K} \label{eq:1bySIRpaLT}
\end{align} 
with (\ref{eq:step1}) and (\ref{eq:1bySIRpaLT}) holding, respectively, because of the IID nature of $\rho_1, \ldots, \rho_K$ and from (\ref{eq:LTrho0}).

\section{Proof of Proposition \ref{prp:SIRUnif_randK}}\label{app:SIRUnif_randK}

Noting that the user's SIR is a valid quantity only for $K\geq1$, the CDF of $\SIRep_k$ with Poisson-distributed $K$ becomes
\begin{align}
F_{\SIRep_k}(\theta) &=  \sum_{\sfk=1}^{\infty} F_{\SIRep_k|K=\sfk}(\theta) \, \frac{f_K(\sfk)}{1-F_K(0)} \label{eq:step00}\\
&= 	\sum_{\sfk=1}^{\ceil{\Na/\theta}-1} \! F_{\SIRep_k|K=\sfk}(\theta) \, \frac{f_K(\sfk)}{1-F_K(0)} + \sum_{\sfk=\ceil{\Na/\theta}}^{\infty} \frac{f_K(\sfk)}{1-F_K(0)} \label{eq:step0}\\
&= 
\frac{1}{e^{\bar{K}}-1} \sum_{\sfk=1}^{\ceil{\Na/\theta}-1} \!\!F_{\SIRep_k|K=\sfk}\!\left(\theta\right) \, \frac{\bar{K}^\sfk}{\sfk!}  + \frac{ 1 -\Gamma \big( \ceil{\Na/\theta},\bar{K} \big)/ \big( \ceil{\Na/\theta}-1 \big)!}{1 - e^{-\bar{K}}} \label{eq:step}
\end{align}
where (\ref{eq:step0}) holds because $F_{\SIRep_k|K=\sfk} (\theta)=1$ for $\theta \geq \Na/\sfk$ while (\ref{eq:step}) follows from the application of (\ref{eq:Poisson PMF})--(\ref{eq:Poisson CDF}). Further applying, in (\ref{eq:step}), the explicit form for $F_{\SIRep_k|K=\sfk} (\cdot)$ from (\ref{eq:CDFSIREP1}) returns (\ref{eq:CDFSIREP_randU0}). 

Alternatively, plugging (\ref{eq:CDFSIREP2}) into (\ref{eq:step}) yields (\ref{eq:CDFSIREP_randU}).

\section{Proof of Propositions \ref{Esterilla} and \ref{Esterilla2}}
\label{Messi}

With a uniform power allocation and a fixed $K$, 
\begin{align}
\bSEep & = \mathbb{E} \! \left[\log_2 \! \left(1+ \SIRep \right)\right] \\
&= \mathbb{E} \! \left[\log_2 \! \left(1+\frac{\Na/K}{1+1/\rho}\right)\right] \\
&= \log_2(e) \, \int_{0}^{\infty} \frac{\mathbb{E} [e^{-z/\rho} ] \left(1 - e^{-z \Na/K} \right)e^{-z}}{z}  \, {d}z \label{eq:usellem1}\\
&=  \log_2(e) \,\int_{0}^{\infty} \frac{ 1-e^{-z \Na/K}}{ {}_1 F_1 \big( 1,1-\delta,z \big)} \, \frac{{d}z}{z}, \label{eq:avSEepfixU} 
\end{align}
where in (\ref{eq:usellem1}) we applied \cite[Lemma 1]{5407601} while (\ref{eq:avSEepfixU}) follows from \cite[Lemma 2]{ICC16-Ganti-Haenggi}.
This proves Prop. \ref{Esterilla}.

Similarly, with an equal-SIR power allocation,
\begin{align}
\bSEpa &= \mathbb{E} \big[\log_2 (1+\SIRpa ) \big] \\
&= \log_2(e) \,\int_{0}^{\infty} \frac{ \mathbb{E} \big[ e^{-z/\SIRpa} \big] \left(1-e^{-z}\right)}{z} \, {d}z \label{eq:usellem2}\\
&= \log_2(e) \,\int_{0}^{\infty} \frac{ 1-e^{-z}}{  {}_1 F_1\left(1,1-\delta,z/\Na \right)^K} \, \frac{{d}z}{z}, \label{eq:bSEpafixU}
\end{align}
where (\ref{eq:usellem2}) follows again from \cite[Lemma 1]{5407601} while (\ref{eq:bSEpafixU}) follows from (\ref{eq:1bySIRpaLT}) in Appendix \ref{app:SIREq_fixedK}. This proves Prop. \ref{Esterilla2}.

\bibliographystyle{IEEEtran}
\bibliography{jour_short,conf_short,references}

\end{document}